\documentclass{IEEEtran}

\usepackage[utf8]{inputenc}

\usepackage{setspace}

\usepackage[colorlinks]{hyperref}

\usepackage{amsmath,amsfonts,amssymb,mathrsfs}
\usepackage{bm}
\usepackage{microtype}
\usepackage{mathtools}

\usepackage{upgreek}

\usepackage{textcomp,nicefrac}

\usepackage{algorithm}
\usepackage{algpseudocode}

\usepackage{glossaries-extra}

\usepackage{graphicx}

\usepackage{caption}
\usepackage[labelformat=simple,font=footnotesize]{subcaption}
\usepackage[dvipsnames]{xcolor}

\usepackage{tikz}
\usetikzlibrary{arrows.meta,shapes,intersections,calc,angles,decorations.pathreplacing,intersections,quotes,angles,positioning,fit,petri,through,shadows,plotmarks,spy}
\usepackage{pgfplots}
\pgfplotsset{compat=1.18}
\usepgfplotslibrary{groupplots}

\usepackage{overpic}

\usepackage{makecell}

\usepackage{booktabs}
\usepackage{multirow}
\usepackage{arydshln}

\usepackage[
	style=ieee,
	dashed=false,
	maxbibnames=20,
	minbibnames=2,
	maxcitenames=1,
	mincitenames=1,
	backend=biber,
	sorting=none,
	useprefix=false
]{biblatex}

\newcommand{\ul}[1]{\underline{#1}}

\newcommand{\textcolormetrics}{white}
\newcommand{\spycolor}{teal}

\newlength{\tempdimw}
\newlength{\tempdimh}
\newlength{\tempdimwb}
\newlength{\tempdimhb}
\newcommand\xspyout{1.575}
\newcommand\yspyout{0.34}
\newcommand\yspyoutaxial{0.34}

\newcommand{\thinmidrule}{%
  \specialrule{0.2pt}{\aboverulesep}{\belowrulesep}%
}

\newcommand{\bolda}{\bm{a}}
\newcommand{\boldb}{\bm{b}}

\newcommand{\boldr}{\bm{r}}
\newcommand{\bolds}{\bm{s}}

\newcommand{\boldx}{\bm{x}}
\newcommand{\boldy}{\bm{y}}
\newcommand{\boldz}{\bm{z}}

\newcommand{\boldlambda}{\bm{\lambda}}
\newcommand{\boldmu}{\bm{\mu}}
\newcommand{\boldtheta}{\bm{\theta}}
\newcommand{\boldphi}{\bm{\phi}}

\newcommand{\boldzero}{\bm{0}}
\newcommand{\boldepsilon}{\bm{\epsilon}}
\newcommand{\boldvarepsilon}{\bm{\varepsilon}}
\newcommand{\boldeta}{\bm{\eta}}

\newcommand{\boldA}{\bm{A}}

\newcommand{\boldE}{\bm{E}}
\newcommand{\boldF}{\bm{F}}

\newcommand{\boldI}{\bm{I}}

\newcommand{\boldR}{\bm{R}}

\newcommand{\calN}{\mathcal{N}}

\newcommand{\calW}{\mathcal{W}}
\newcommand{\calX}{\mathcal{X}}
\newcommand{\calY}{\mathcal{Y}}
\newcommand{\calZ}{\mathcal{Z}}

\newcommand{\rma}{\mathrm{a}}

\newcommand{\rme}{\mathrm{e}}

\newcommand{\rmh}{\mathrm{h}}

\newcommand{\rml}{\mathrm{l}}

\newcommand{\rmp}{\mathrm{p}}

\newcommand{\rmt}{\mathrm{t}}

\newcommand{\ybar}{\bar{y}}
\newcommand{\boldybar}{\bar{\boldy}}

\newcommand{\boldxhat}{\hat{\boldx}}
\newcommand{\boldzhat}{\hat{\boldz}}

\newcommand{\R}{\mathbb{R}}

\newcommand{\transp}{^\top}

\pgfplotsset{
	ref/.style={
		color=NavyBlue, mark=*, mark size=1pt,
	},
	ref_line/.style={
		ref, no marks,
	},
	ref_points/.style={
		ref, only marks,
	},
}

\pgfplotsset{
	mlaa/.style={
		color=BurntOrange, mark=square*, mark size=1pt,
	},
	mlaa_line/.style={
		mlaa, no marks,
	},
	mlaa_points/.style={
		mlaa, only marks,
	},
}

\pgfplotsset{
	unet/.style={
		color=ForestGreen, mark=triangle*, mark size=1pt,
	},
	unet_line/.style={
		unet, no marks,
	},
	unet_points/.style={
		unet, only marks,
	},
}

\pgfplotsset{
	dpstof/.style={
		color=Red, mark=+, mark size=1pt,
	},
}

\pgfplotsset{
	ref/.style={
		color=NavyBlue, mark=*, mark size=1pt,
	},
	ref_line/.style={
		ref, no marks,
	},
	ref_points/.style={
		ref, only marks,
	},
}

\pgfplotsset{
	mlaa/.style={
		color=BurntOrange, mark=square*, mark size=1pt,
	},
	mlaa_line/.style={
		mlaa, no marks,
	},
	mlaa_points/.style={
		mlaa, only marks,
	},
}

\pgfplotsset{
	unet/.style={
		color=ForestGreen, mark=triangle*, mark size=1pt,
	},
	unet_line/.style={
		unet, no marks,
	},
	unet_points/.style={
		unet, only marks,
	},
}

\pgfplotsset{
	dpstof/.style={
		color=Red, mark=+, mark size=1pt,
	},
	dpstof_line/.style={
		dpstof, no marks,
	},
	dpstof_points/.style={
		dpstof, only marks,
	},
}

\pgfplotsset{
	dpsnotof/.style={
		color=Orchid, mark=x, mark size=1pt,
	},
	dpsnotof_line/.style={
		dpsnotof, no marks,
	},
	dpsnotof_points/.style={
		dpsnotof, only marks,
	},
}

\pgfplotsset{
	dpstofosem/.style={
		color=RawSienna, mark=+, mark size=1pt,
	},
	dpstofosem_line/.style={
		dpstofosem, no marks,
	},
	dpstofosem_points/.style={
		dpstofosem, only marks,
	},
	dpstof_line/.style={
		dpstof, no marks,
	},
	dpstof_points/.style={
		dpstof, only marks,
	},
}

\pgfplotsset{
	dpsnotof/.style={
		color=Orchid, mark=x, mark size=1pt,
	},
	dpsnotof_line/.style={
		dpsnotof, no marks,
	},
	dpsnotof_points/.style={
		dpsnotof, only marks,
	},
}

\pgfplotsset{
	dpstofosem/.style={
		color=RawSienna, mark=+, mark size=1pt,
	},
	dpstofosem_line/.style={
		dpstofosem, no marks,
	},
	dpstofosem_points/.style={
		dpstofosem, only marks,
	},
}

\setabbreviationstyle{long-short}
\glsdisablehyper
\newabbreviation{2D}{2-D}{two-dimensional}
\newabbreviation{3D}{3-D}{three-dimensional}
\newabbreviation{dD}{$d$-D}{$d$-dimensional}
\newabbreviation{mD}{$m$-D}{$m$-dimensional}
\newabbreviation{AC}{AC}{attenuation correction}
\newabbreviation{ADMM}{ADMM}{alternating direction method of multipliers}
\newabbreviation{ALARA}{ALARA}{As Low As Reasonnably Achievable}
\newabbreviation{CAOL}{CAOL}{convolutional analysis operator learning}
\newabbreviation{CDL}{CDL}{convolutional DiL}
\newabbreviation{CNN}{CNN}{convolutional neural network}
\newabbreviation{CRC}{CRC}{contrast recovery coefficient}
\newabbreviation{CS}{CS}{compressive sensing}
\newabbreviation{CT}{CT}{computed tomography}
\newabbreviation{DDIM}{DDIM}{denoising diffusion implicit model}
\newabbreviation{DDPM}{DDPM}{denoising diffusion probabilistic model}
\newabbreviation{DECT}{DECT}{dual-energy computed tomography}
\newabbreviation{DiL}{DiL}{dictionary learning}
\newabbreviation{DL}{DL}{deep learning}
\newabbreviation{AI}{AI}{artificial intelligence}
\newabbreviation{DM}{DM}{diffusion model}
\newabbreviation{DPS}{DPS}{diffusion posterior sampling}
\newabbreviation{SDE}{SDE}{stochastic differential equation}
\newabbreviation{DTV}{DTV}{directional total variation}
\newabbreviation{DWT}{DWT}{discrete wavelet transform}
\newabbreviation{EM}{EM}{expectation-maximization}
\newabbreviation{FBP}{FBP}{filtered backprojection}
\newabbreviation{FC}{FC}{full-count}
\newabbreviation{FWHM}{FWHM}{full width at half maximum}
\newabbreviation{GAN}{GAN}{generative adversarial network}
\newabbreviation{GT}{GT}{ground truth}
\newabbreviation{HC}{HC}{high-count}
\newabbreviation{HU}{HU}{Houndsfield Units}
\newabbreviation{FOV}{FOV}{field of view}
\newabbreviation{IDWT}{IDWT}{inverse discrete wavelet transform}
\newabbreviation{IFFT}{IFFT}{inverse fast Fourier transform}
\newabbreviation{JRAA}{JRAA}{joint reconstruction of activity and attenuation}
\newabbreviation{JTV}{JTV}{joint total variation}
\newabbreviation{KL}{KL}{Kullback-Leibler}
\newabbreviation{LC}{LC}{low-count}
\newabbreviation{LAC}{LAC}{linear attenuation coefficient}
\newabbreviation{LDM}{LDM}{latent diffusion model}
\newabbreviation{LBFGS}{L-BFGS}{limited-memory  Broyden-Fletcher-Goldfarb-Shanno}
\newabbreviation[
	longplural={lines of response},
	firstplural={lines of response (LORs)}
]{LOR}{LOR}{line of response}
\newabbreviation{MAP}{MAP}{maximum \emph{a~posteriori}}
\newabbreviation{MAPEM}{MAPEM}{maximum \textit{a posteriori} expectation maximization}
\newabbreviation{MBIR}{MBIR}{model-based iterative reconstruction}
\newabbreviation{MCDL}{CDL}{multichannel convolutional dictionary learning}
\newabbreviation{MCAOL}{MCAOL}{multichannel convolutional analysis operator learning}
\newabbreviation{ML}{ML}{maximum likelihood}
\newabbreviation{MLAA}{MLAA}{maximum likelihood activity and attenuation}
\newabbreviation{ACF}{ACF}{attenuation correction factor}
\newabbreviation{MLACF}{MLACF}{maximum likelihood activity and attenuation correction factors}
\newabbreviation{MLEM}{MLEM}{maximum-likelihood expectation-maximization}
\newabbreviation{MNIST}{MNIST}{Modified National Institute of Standards and Technology}
\newabbreviation{MR}{MR}{magnetic resonance}
\newabbreviation{MRI}{MRI}{magnetic resonance imaging}
\newabbreviation{MSE}{MSE}{mean squared error}
\newabbreviation{VOI}{VOI}{volume of interest}
\newabbreviation{MVAE}{MVAE}{multi-branch VAE}
\newabbreviation{NAC}{NAC}{non attenuation-corrected}
\newabbreviation{NN}{NN}{neural network}
\newabbreviation{NRMSE}{NRMSE}{normalized root mean squared error}
\newabbreviation{OOD}{OOD}{out-of-distribution}
\newabbreviation{OSEM}{OSEM}{ordered subsets expectation maximization}
\newabbreviation{PCA}{PCA}{principal component analysis}
\newabbreviation{PCCT}{PCCT}{photon-counting computed tomography}
\newabbreviation{PDF}{PDF}{probability distribution function}
\newabbreviation{PET}{PET}{positron emission tomography}
\newabbreviation{PLS}{PLS}{parallel level set}
\newabbreviation{PML}{PML}{penalized maximum likelihood}
\newabbreviation{PVE}{PVE}{partial volume effect}
\newabbreviation{PSNR}{PSNR}{peak signal-to-noise ratio}
\newabbreviation{PWLS}{PWLS}{penalized weighted least squares}
\newabbreviation{RC}{RC}{recovery coefficient}
\newabbreviation{ReLU}{RelU}{rectified linear unit}
\newabbreviation{ROI}{ROI}{region of interest}
\newabbreviation{SPECT}{SPECT}{single-photon emission CT}
\newabbreviation{SPS}{SPS}{separable paraboloidal surrogates}
\newabbreviation{SQS}{SQS}{separable quadratic surrogate}
\newabbreviation{SNR}{SNR}{signal-to-noise ratio}
\newabbreviation{SSIM}{SSIM}{structural similarity index measure}
\newabbreviation{SSS}{SSS}{single scatter simulation}
\newabbreviation{TNV}{TNV}{total nuclear variation}
\newabbreviation{TOF}{TOF}{time-of-flight}
\newabbreviation{TV}{TV}{total variation}
\newabbreviation{VAE}{VAE}{variational autoencoder}
\newabbreviation{beta-VAE}{$\beta$-VAE}{beta-variational autoencoder}
\newabbreviation{WDM}{WDM}{wavelet diffusion model}
\newabbreviation{WGAN}{W-GAN}{Wasserstein GAN}
\newabbreviation{WLS}{WLS}{weighted least squares}
\newabbreviation{STD}{STD}{standard deviation}
\newabbreviation{SF}{SF}{scatter fraction}
\newabbreviation{XCAT}{XCAT}{extended cardiac-torso}
\newabbreviation{FDG}{\textsuperscript{18}F-FDG}{[\textsuperscript{18}F]Fluorodeoxyglucose}
\newabbreviation{PSMA}{[\textsuperscript{18}F]-PSMA}{[\textsuperscript{18}F]-\emph{prostate-specific membrane antigen-11}}

\begin{document}
	
	\title{Joint Reconstruction of Activity and Attenuation in PET by Diffusion Posterior Sampling in Wavelet Coefficient Space} 
	
	\author{
		Clémentine Phung-Ngoc, Alexandre Bousse, Antoine De Paepe, Thibaut Merlin, Baptiste Laurent, Hong-Phuong~Dang, Olivier Saut, Catherine Cheze-Le-Rest and Dimitris Visvikis, \IEEEmembership{Fellow, IEEE}
		\thanks{
			This work involved human subjects or animals in its research. The authors confirm that all human/animal subject research procedures and protocols are exempt from review board approval.
		}
		\thanks{
			This work was supported by the French National Institute of Health and Medical Research (Inserm), the French Institute for Research in Computer Science and Automation (Inria), the French National Research Agency (ANR) under grant No ANR-20-CE45-0020 and by France Life Imaging under grant No ANR-11-INBS-0006.
			}
		\thanks{
			Clémentine Phung-Ngoc, Alexandre Bousse, Antoine De Paepe, Thibaut Merlin, Baptiste Laurent, Catherine Cheze-Le-Rest and Dimitris Visvikis are with LaTIM, Inserm UMR 1101, Université de Bretagne Occidentale, 29238 Brest, France.
			}
		\thanks{
			Hong-Phuong Dang is with CentraleSupélec, IETR, CNRS UMR 6164, 35576 Cesson-Sévigné, France.
			}
		\thanks{
			Olivier Saut is with Inria Monc, Université de Bordeaux, Bordeaux INP, CNRS, 33405 Talence, France.
			}
		\thanks{
			Catherine Cheze-Le-Rest is also with Nuclear Medicine Department, Poitiers University Hospital, F-86022, Poitiers, France.
			}
		\thanks{
			Corresponding author: A. Bousse; email: \href{mailto:bousse@univ-brest.fr}{\texttt{bousse@univ-brest.fr}}.
			}
	}
	
	\maketitle
	
	\begin{abstract}
	
	\Gls{AC} is necessary for accurate activity quantification in \gls{PET}. Conventional reconstruction methods typically rely on attenuation maps derived from a co-registered \gls{CT} or \gls{MR} scan. However, this additional scan may complicate the imaging workflow, introduce misalignment artifacts and increase radiation exposure. In this paper, we propose a \gls{JRAA} approach that eliminates the need for auxiliary anatomical imaging by relying solely on emission data. This framework combines \gls{WDM} and \gls{DPS} to reconstruct fully \gls{3D} data. Experimental results on simulated data show our method outperforms \gls{MLAA} and \gls{MLAA}-UNet with U-Net-based postprocessing, and yields high-quality noise-free reconstructions across various count settings with \gls{TOF}.  
	It is also able to reconstruct non-\gls{TOF} data, although the reconstruction quality significantly degrades in \gls{LC} conditions, limiting its practical effectiveness in such settings. Nonetheless, a non-\gls{TOF} Biograph mMR real data reconstruction with joint scatter estimation highlights the potential of the method for clinical applications. This approach represents a step towards stand-alone \gls{PET} imaging by reducing the dependence on anatomical modalities while maintaining quantification accuracy, even in \gls{LC} scenarios when \gls{TOF} information is available. Our code is available on GitHub at \href{https://github.com/clemphg/jraa-dps}{\texttt{https://github.com/clemphg/jraa-dps}}.
\end{abstract}

\glsresetall

	\glsresetall

	
	\vspace{-0.2cm}

\section{Introduction}

\glsunset{PET} 

\IEEEPARstart{P}{ositron} emission tomography (PET) is a nuclear medical imaging technique used to visualize functional mechanisms within the body. It involves injecting a positron-emitting radiopharmaceutical as a tracer, designed specifically for the targeted function. This tracer concentrates in the regions of interest via the metabolism, and pairs of $\upgamma$ photons resulting from the  electron-positron annihilation events are detected by an array of ring detectors placed around the patient. Finally, an image of the activity distribution is reconstructed from the emission data. \Gls{PET} has a wide range of applications in fields such as oncology, cardiology or neurology. 

\Gls{PET} technology has undergone significant advancements since its introduction in the 1970s~\cite{jones2017history}, particularly in terms of sensitivity, spatial resolution, and quantitative accuracy. Modern \gls{PET} systems are typically integrated with anatomical imaging modalities such as \gls{CT} or \gls{MR}, enabling the generation of $\upgamma$ photon attenuation maps for \gls{AC}\footnote{In the following, the acronym ``AC'' stands for ``attenuation correction'' and for ``attenuation-corrected'' (depending on the phrasing).} and scatter estimation. These hybrid systems not only facilitate more accurate quantitative imaging but also provide complementary functional and anatomical information, thereby assisting clinicians in diagnosis. Furthermore, contemporary \gls{PET} systems increasingly leverage \gls{TOF} information to enhance spatial localization.

In recent years, several researchers have explored the possibility of performing \gls{JRAA} using emission data alone. \citeauthor{defrise2012time}~\cite{defrise2012time} demonstrated that, in \gls{TOF} \gls{PET}, it is theoretically feasible to estimate both the activity distribution and the attenuation coefficients up to a multiplicative constant, provided the \gls{TOF} resolution is sufficiently high. Building on this foundation, numerous \gls{JRAA} methods have been proposed with promising results, including \gls{MLAA}~\cite{rezaei2012simultaneous} and other similar methods \cite{rezaei2014ml} (see \citeauthor{berker2016attenuation}~\cite{berker2016attenuation} for a review). These advancements open new avenues for improving \gls{PET} imaging, such as reducing patient radiation dose, which could enable more frequent scans for improved monitoring and patient screening. They also offer opportunities for enhanced respiratory motion management, as the reliance on \gls{CT} snapshots at each respiratory phase could be mitigated~\cite{elhamiasl2025joint}.   

The above-mentioned \gls{JRAA} techniques are not without limitations. In particular, they suffer from activity--attenuation crosstalk when the \gls{TOF} resolution is insufficient, and they fail entirely in the absence of \gls{TOF} information. 

\Gls{AI}  and more particularly \gls{DL} techniques have changed the paradigm in \gls{PET} imaging, for image  reconstruction \cite{reader2020deep} and post-processing \cite{bousse2024review}. \Gls{DL} has also been investigated for \gls{AC} in \gls{PET}, where approaches for \gls{PET} \gls{AC} from emission data only can be divided in two main categories \cite{chen2023deep}. 

Direct methods use deep models to directly estimate the \gls{AC} activity image from the non-\gls{AC} image \cite{shiri2019direct, dong2020deep}. They are particularly sensitive to the characteristics of the training data, since they learn a direct image-to-image mapping. In such approaches, the measured emission data are not explicitly used to guide the reconstruction once the non-\gls{AC} image has been formed. Although these methods can provide good results when applied to data that are well represented in the training set, they do not enforce consistency with the measured \gls{PET} data and may therefore be less robust in out-of-distribution cases, for example in the presence of unusual anatomy, lesions, or different count levels.

Indirect \gls{AC} methods aim to predict a synthetic attenuation map, which can then be incorporated into a conventional iterative reconstruction algorithm such as \gls{EM}. As a result, the final activity reconstruction remains guided by the emission measurements. The attenuation map can be predicted from the non-\gls{AC} \gls{PET} image \cite{liu2018deep,dong2019synthetic}, or from an \gls{MLAA} reconstruction \cite{hwang2018improving}. However, both strategies still depend on intermediate reconstructions that are themselves noisy, and the quality of the final attenuation map remains strongly linked to the acquisition settings represented in the training data.

One way to mitigate this dependence on acquisition settings is to use diffusion-prior-based \gls{JRAA} methods. \Glspl{DM} have arisen as an alternative to \glspl{GAN} for image generation and other tasks \cite{dhariwal2021diffusion}, and have been widely explored in medical imaging \cite{kazerouni2023diffusion}. Score-based \glspl{DM} are a generative technique which consists in gradually transforming a collection of images into white noise via a \gls{SDE}, then, in learning the inverse \gls{SDE} by score matching using a \gls{CNN}, and finally to generate images from random white noise. The trained \glspl{DM} can be used to solve inverse problems with approaches such as \gls{DPS} and have been used in image reconstruction such as in \gls{PET} \cite{singh2024score,webber2025likelihood} and \gls{CT} \cite{li2024ct,vazia2025material}.  In this setting, the \gls{DM} acts as a prior in the reconstruction algorithm rather than as a direct supervised predictor. Consequently, acquisition-specific factors, including scan duration, noise level, \gls{TOF} resolution, and scanner geometry, are explicitly accounted for in the forward model and data-consistency term. This makes the approach less dependent on the specific acquisition conditions represented in the training set, while ensuring that the reconstructed images remain consistent with the measured emission data.

In this work, we propose to leverage \glspl{DM} for the task of \gls{JRAA} via \gls{DPS}, using a model trained on activity--attenuation image pair. The proposed \gls{JRAA}-\gls{DPS} method (cf. Figure~\ref{fig:schema_dps}) extends our preliminary study \cite{phung2024joint}, which was limited to \gls{2D} phantoms. \citeauthor{bae2025joint}~\cite{bae2025joint} presented a similar approach, although their \gls{DM} operates on attenuation images only.

The main contributions of this work are as follows:
\begin{itemize}
	\item Building on our previous study, we propose a \gls{DM}-based \gls{JRAA} method, namely \gls{JRAA}-\gls{DPS}, which leverages a diffusion prior learned from activity--attenuation pairs for the joint reconstruction of activity and attenuation from emission data alone.
	\item The \gls{DM} prior itself is decoupled from the acquisition model. Consequently, the same trained model can be applied across different acquisition settings, including scan duration, \gls{TOF} resolution and scanner geometry, without retraining, provided that the forward model is adapted to the measured data.
	\item To process large \gls{3D} volumes, we employed a \gls{WDM} trained on eight-channel wavelet coefficient images \cite{friedrich2024wdm}.
\end{itemize}

The rest of this paper is organized as follows. Section~\ref{sec:background} provides background on \gls{JRAA}. Section~\ref{sec:method} describes our proposed method, namely \gls{JRAA}-\gls{DPS}, followed by details of the experimental setup on simulated and real data in Section~\ref{sec:experiments}. Section~\ref{sec:results} presents the experimental results, and Section~\ref{sec:discussion} discusses the limitations of \gls{JRAA}-\gls{DPS}. Finally, Section~\ref{sec:conclusion} concludes the paper.


\subsection*{Nomenclature}

`$\transp$' denotes the matrix transposition. For a real-valued vector $\bolda$, $[\bolda]_i$ denotes its $i$th entry, while for a real-valued matrix $\boldA$, $[\boldA]_{i,j}$ denotes the entry in the $i$th row and $j$th column. For a given real finite-dimensional vector space $\calZ$, $\boldzero_{\calZ}$ and $\boldI_{\calZ}$ and respectively denote the null vector and identity matrix on $\calZ$.

$\calX \triangleq \R^{D \times{}W\times{}H} \cong   \R^{m}$ and $\calY \triangleq \R^{n_{\rml}\times n_{\rmt}}$ respectively denote the image and measurement spaces, where $D$, $W$ and $H$ are respectively the depth, width and height of the image, $m=DWH$ is the number of voxels, $n_{\rml}$ is the number of \glspl{LOR} and $n_{\rmt}$ is the number of \gls{TOF} bins ($n_{\rmt}=1$ in non-\gls{TOF} \gls{PET}). The activity and attenuation images are respectively denoted by the column vectors
\begin{displaymath}
	\boldlambda \triangleq [\lambda_1,\dots,\lambda_m]\transp \in \calX
	\quad\text{and}\quad
	\boldmu \triangleq [\mu_1,\dots,\mu_m]\transp \in \calX,
\end{displaymath}
where $\lambda_j$ and $\mu_j$ respectively denote the activity and attenuation at voxel $j \in \{1,\dots,m\}$. The measurement, denoted $\boldy$, is given by a matrix
\begin{displaymath}
\boldy = 
\begin{bmatrix} 
	y_{1,1} & \dots  & y_{1,n_{\rmt}}\\
	\vdots & \ddots & \vdots\\
	y_{n_{\rml},1} & \dots  & y_{n_{\rml},n_{\rmt}} 
\end{bmatrix} \in \calY
\end{displaymath}
such that $y_{i,k} \triangleq [\boldy]_{i,k}$ is the number of detected coincidences at the $i$th \gls{LOR} and the $k$th \gls{TOF} bin.

	\section{Background on Joint Activity and Attenuation Reconstruction}\label{sec:background}

\begin{figure*}[htbp]
    \centering

    \includegraphics[width=0.95\textwidth]{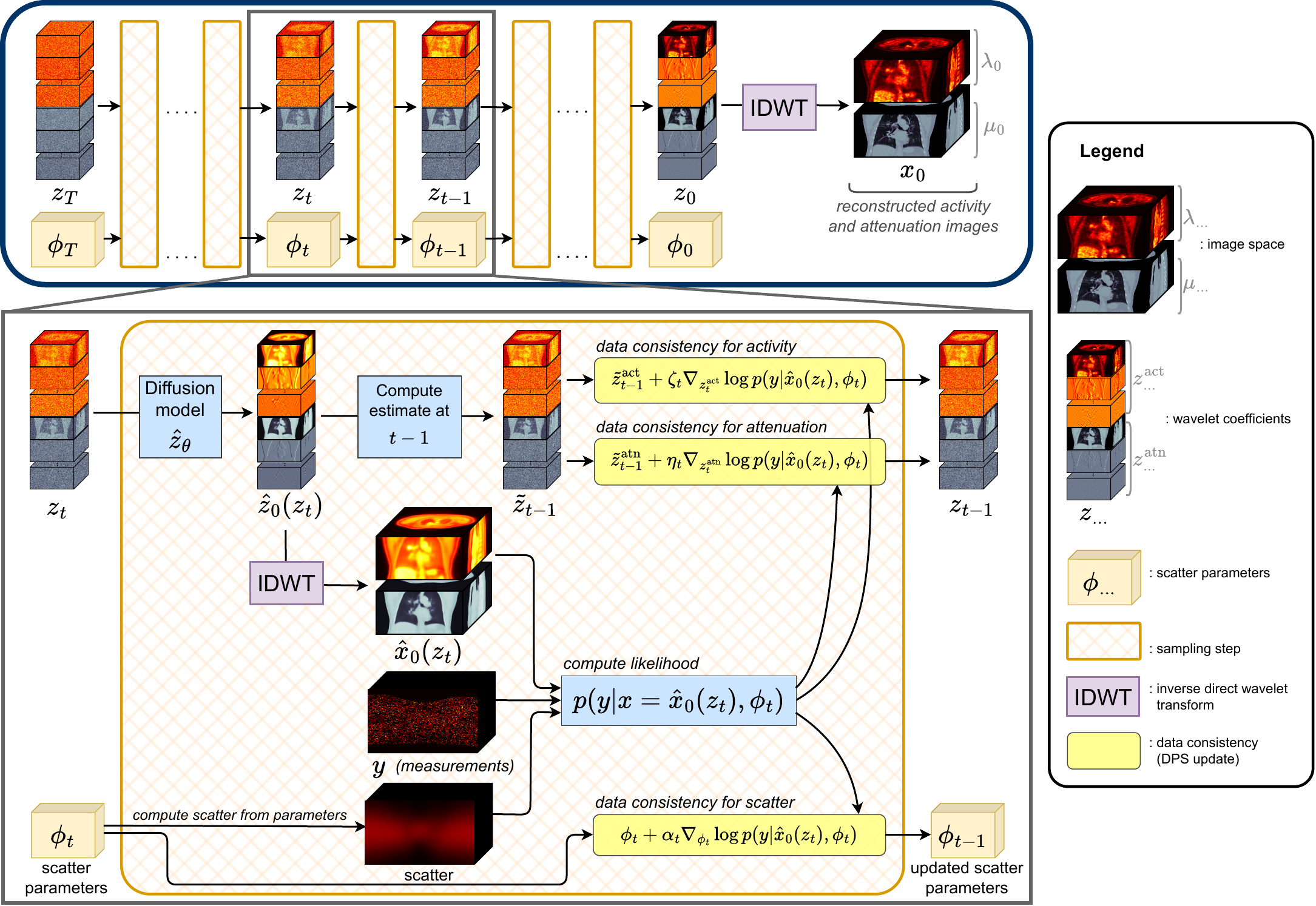}

    \caption{Proposed reconstruction framework for \gls{JRAA}-\gls{DPS}. A \gls{WDM} is trained to generate activity-attenuation image pairs and \gls{DPS} is employed to reconstruct from measurements. Joint scatter estimation can be integrated into the framework. Bypassing data consistency (i.e., \gls{DPS} updates) generates random samples from the learned prior distribution.}
    \label{fig:schema_dps}
\end{figure*}


\subsection{PET Reconstruction with Known Attenuation}

In \gls{PET} imaging the task is to reconstruct a radioactivity distribution image $\boldlambda \in \calX$, lying in an attenuation medium $\boldmu \in \calX$, from a measurement $\boldy \in \calY$. 

For each \gls{LOR} $i$ and each \gls{TOF} bin $k$, the measurement $y_{i,k}$ is a random variable with conditional expectation $\ybar_{i,k}(\boldlambda,\boldmu) \triangleq \mathbb{E}[y_{i,k} |  \boldlambda,\boldmu]$ given  $(\boldlambda,\boldmu)$ and following a Poisson distribution, i.e.,
\begin{equation}\label{eq:poisson}
	(y_{i,k} \mid \boldlambda, \boldmu) \sim \mathrm{Poisson} \big( \ybar_{i,k} (\boldlambda,\boldmu) \big) \, ,
\end{equation} 
with independent entries $y_{i,k}$. The expected measurement is given by
\begin{align}\label{eq:ybar}
	\ybar_{i,k} (\boldlambda,\boldmu) & = \tau \cdot a_i(\boldmu) \cdot \sum_{j=1}^m p_{i,k,j} \lambda_j  + r_{i,k} + s_{i,k} 
\end{align} 
where $\tau$ is the scan duration,  $p_{i,k,j}$ is the system's detection probability that an emission from voxel $j$ is detected in $(i,k)$ considering geometry, resolution and sensitivity, while $r_{i,k}$ and $s_{i,k}$ are respectively the expected contribution of randoms and scatter. The attenuation coefficient $a_i(\boldmu)$ is given by the Beer-Lambert law as
\begin{equation}\label{eq:attn}
	a_i(\boldmu)  = \rme^{- [\boldR \boldmu]_i } \, ,
\end{equation}
where $\boldR  \in \R^{n_{\rml} \times m }$ is the discrete Radon transform which performs line integrals along each \gls{LOR} $i=1,\dots,n_{\rml}$. Furthermore, we denote the forward model  by $\boldybar \colon \calX \times \calX \to \calY$ which is obtained by regrouping the $\ybar_{i,k}$s in a matrix with same shape as $\boldy$.

When the attenuation map $\boldmu$ is known, for example derived from a co-registered anatomical image such as \gls{CT} or \gls{MR}, reconstructing $\boldlambda$ can be achieved by solving the \gls{MAP} optimization problem
\begin{equation}\label{eq:map}
	\max_{\boldlambda} \, p(\boldy \mid \boldlambda,\boldmu) \cdot p(\boldlambda)
\end{equation}
where the conditional \gls{PDF} $p(\boldy | \boldlambda,\boldmu) $ is given by \eqref{eq:poisson} and \eqref{eq:ybar}. In general, the prior \gls{PDF} $p(\boldlambda)$ is unknown and is replaced by a function of the form $p(\boldlambda) \propto \exp(-\beta R(\boldlambda))$ where $\beta>0$ is a weight and $R~ ~\colon~\calX~\to~\R$ is a convex regularizer that promotes image smoothness while preserving edges (see for example \citeauthor{nuyts2002concave}~\cite{nuyts2002concave}). In practice, solving \eqref{eq:map} is achieved in a post-log form by 
\begin{equation}\label{eq:pml}
	\min_{\boldlambda} \, \ell\big( \boldy , \boldybar(\boldlambda,\boldmu)  \big) +  \beta R(\boldlambda)
\end{equation}
where for the Poisson model $- \log p(\boldy \mid \boldlambda,\boldmu) = \ell( \boldy , \boldybar(\boldlambda,\boldmu)  ) $ with $\ell(\boldy,\boldybar ) \triangleq \sum_{i,k} -y_{i,k} \log \ybar_{i,k} + \ybar_{i,k}$ (omitting terms that are independent of $\boldlambda$). Solving \eqref{eq:pml} can be achieved with iterative algorithms such as the \gls{EM} algorithm \cite{shepp1982maximum} ($\beta=0$) or its penalized version \cite{depierro1995} ($\beta>0$).


\subsection{Joint PET Activity and Attenuation Reconstruction}

In the absence of \gls{CT} or \gls{MR}, the attenuation image $\boldmu$ is most likely unknown and therefore solving \eqref{eq:pml} is impossible by \gls{EM}. Alternatively, \gls{JRAA} can be achieved by solving the \gls{MAP} optimization problem in $\boldx = (\boldlambda,\boldmu)$,
\begin{equation}\label{eq:map_aa}
	\max_{\boldx} \, p(\boldy \mid \boldx) \cdot p(\boldx) \, ,
\end{equation}
where $p(\boldx)$ is the joint \gls{PDF} of $(\boldlambda,\boldmu)$, which can be rewritten in a post-log form as 
\begin{equation}\label{eq:pml_aa}
	\min_{\boldx} \, \ell\big( \boldy , \boldybar(\boldx)  \big) +  \beta R(\boldx)
\end{equation}
where $R(\boldx)$ is a regularizer such that $p(\boldx) \propto \exp(-\beta R(\boldlambda,\boldmu))$ and that acts on $\boldlambda$ and $\boldmu$ separately or synergistically by exploiting shared information between the two images (see for example \citeauthor{arridge2021overview}~\cite{arridge2021overview}).
Solving \eqref{eq:pml_aa} can be achieved with  an iterative algorithm  alternating between minimization in $\boldlambda$ and $\boldmu$. When $\beta=0$, this corresponds to the \gls{MLAA} method \cite{rezaei2012simultaneous}. An alternative approach consists in estimating the \glspl{ACF} $a_i(\boldmu)$ instead of the image $\boldmu$ \cite{defrise2012time,rezaei2014ml}. 

However, \gls{MLAA} suffers from $\boldlambda$--$\boldmu$ crosstalk when the \gls{TOF} resolution is too low, and fails to reconstruct useful images in the absence of \gls{TOF} information. Additionally,  \gls{MLAA} suffers from scaling issues that need to be addressed specifically (see for example \citeauthor{li2020practical}~\cite{li2020practical}). 

	\section{Proposed Approach: Joint Reconstruction of the Activity and the Attenuation by Diffusion Posterior Sampling}\label{sec:method}

We propose to use \glspl{DM} and \gls{DPS} to leverage joint prior \gls{PDF} $p(\boldlambda,\boldmu)$ to address activity--attenuation crosstalk and scaling factor issues in \gls{JRAA}. The proposed framework, namely \gls{JRAA}-\gls{DPS}, is summarized in Figure~\ref{fig:schema_dps}.


\subsection{Diffusion Models}\label{sec:dm}

In the following, $\boldx$ denotes a multichannel image which lies in the vector space $\calW = \calX^d$, where $d$ is the number of channels (in the case of \gls{JRAA}, $d=2$ and $\boldx = (\boldlambda,\boldmu)$). The measurement  $\boldy\in\calY$ follows a conditional \gls{PDF} $p(\boldy|\boldx)$ given $\boldx$.

\Glspl{DM} are generative models made of two distinct components: the forward process and reverse process (i.e., sampling).  

During the forward process, an image $\boldx = \boldx_{0}$ is sampled from the training dataset with a \gls{PDF} $p^\mathrm{data}(\boldx)$ and is then gradually transformed into white noise over $T$ steps. A \gls{CNN} is then trained to predict the score in order to estimate the noise-free image $\boldx_{0}$ from its noisy version  $\boldx_t$ at each time $t\in \{1, \dots, T\}$ (reverse process).
In \glspl{DDPM} \cite{ho2020denoising}, $\boldx_t$ is sampled from $\boldx_{t-1}$ as
\begin{equation}\label{eq:ddpm-sample}
	\boldx_{t} =  \sqrt{\alpha_{t}} \boldx_{t-1} +  \sqrt{1-\alpha_{t}} \boldepsilon_t\, ,\quad \boldepsilon_t \sim \mathcal{N}(\boldzero_{\calW} ,  \boldI_{\calW})
\end{equation}
where the $\boldepsilon_t$ are independent and $\alpha_t$ is a decreasing scaling factor defined such that $\boldx_T \sim \calN (\boldzero_{\calW}, \boldI_{\calW})$. Thus, the sequence $\boldx_0,\boldx_1,\boldx_2,\dots$ is a Markov chain.

The reverse process consists in sampling  $\boldx_{t-1}$ from $\boldx_{t}$, starting from $\boldx_{T} \sim \mathcal{N}(\boldzero_{\calW} ,  \boldI_{\calW})$, which is not possible as $p(\boldx_{t-1}|\boldx_{t})$ is unknown. Nevertheless, the conditional \gls{PDF} of $\boldx_{t-1}$ given $(\boldx_0,\boldx_{t})$ is known \cite{ho2020denoising}:
\begin{align}
	\boldx_{t-1} & = \frac{\sqrt{\alpha_t} (1-\bar{\alpha}_{t-1})}{1-\bar{\alpha}_t} \boldx_t + \frac{\sqrt{\bar{\alpha}_{t-1}} \beta_t}{1 - \bar{\alpha}_t} \boldx_0 + \sigma_t \boldvarepsilon_t  \label{eq:ddpm-reverse1}\\
	\boldvarepsilon_t & \sim \calN(\boldzero_{\calW},\boldI_{\calW}) \, , \nonumber
\end{align}
where $\bar{\alpha}_t = \prod_{s=1}^t \alpha_s$, $\beta_t = 1-\alpha_t$, $\sigma_t^2= \beta_t(1-\bar{\alpha}_{t-1})/(1-\bar{\alpha}_{t})$.  To effectively  sample $\boldx_{t-1}$ from $\boldx_{t}$, \eqref{eq:ddpm-reverse1} is approximated by replacing $\boldx_0$ with its conditional expectancy given $\boldx_t$, denoted by  $\boldxhat_0(\boldx_t) \triangleq \mathbb{E}[\boldx_0 | \boldx_t]$, which is known through Tweedie's formula:
\begin{equation}\label{eq:x0}
	\boldxhat_0 (\boldx_t) = \frac{1}{\sqrt{\bar{\alpha}_t}} \big(\boldx_t + (1-\bar{\alpha}_t) \nabla  \log p_t(\boldx_t)\big)\, ,
\end{equation}
where $p_t$ is the \gls{PDF} of $\boldx_t$. 

The mapping $\boldx_t \mapsto \nabla  \log p_t(\boldx_t)$, is unknown, and is therefore approximated by a \gls{CNN} $\bolds_{\boldtheta} \colon \calW \times [0, T] \to \calW$ parametrized with $\boldtheta$, which can be trained via score matching \cite{song2019generative}  as 
\begin{equation}\label{eq:score_matching}
	\min_{\boldtheta} \, \mathbb{E}_{t, \boldx_0, \boldepsilon} \left[ \big\| \bolds_{\boldtheta}(\boldx_t,t) - \nabla_{\boldx_t} \log p_t(\boldx_t \mid \boldx_0) \big\|_2^2 \right]
\end{equation}
where $\boldx_0 \sim p^\mathrm{data}$, $\boldx_t = \sqrt{\bar{\alpha}_t} \boldx_0 + \sqrt{1-\bar{\alpha}_t}\boldepsilon$ with $\boldepsilon~\sim~\calN (\boldzero_{\calW}, \boldI_{\calW})$, and $\log p_t(\boldx_t | \boldx_0)$ is known by telescoping \eqref{eq:ddpm-reverse1}. An equivalent formulation involves a denoising \gls{CNN} $\boldxhat_{\boldtheta} \colon \calW \times [0, T] \to \calW$, which relates to $\bolds_{\boldtheta}$ through 
\begin{equation}\label{eq:x0nn}
	\boldxhat_{\boldtheta} (\boldx_t,t) = \frac{1}{\sqrt{\bar{\alpha}_t}} \left(\boldx_t + (1-\bar{\alpha}_t) \bolds_{\boldtheta}(\boldx_t,t) \right)\, ,
\end{equation}
and is trained to approximate the current denoised image as
\begin{equation}\label{eq:score_matching-eq}
	\min_{\boldtheta} \, \mathbb{E}_{t, \boldx_0, \boldepsilon} \left[ \left\| \boldxhat_{\boldtheta}(\boldx_t,t) - \boldx_0 \right\|_2^2 \right]\, ,
\end{equation}
which can then be used to sample $\boldx_{t-1}$ from  $\boldx_{t}$ by replacing $\boldx_0$ with $\boldxhat_{\boldtheta}(\boldx_t,t)$ in \eqref{eq:ddpm-reverse1}. 

Thus, a sequence of multichannel images $\boldx_{T-1},\boldx_{T-2},\dots$ can be sampled, starting from $\boldx_{T} \sim \calN(\boldzero_{\calW},\boldI_{\calW})$, such that the final image $\boldx_0$ is sampled with a \gls{PDF} that is a generalization of $p^\mathrm{data}$ (through the \gls{CNN} approximation) and that mimics the true prior  $p(\boldx)$.

\Glspl{DDIM}~\cite{song2020denoising} reuse the same training objective as \glspl{DDPM} but replace the stochastic reverse diffusion \eqref{eq:ddpm-reverse1} with a deterministic procedure. This update does not contain a stochastic noise term and therefore deterministically transports $\boldx_t$ along a straight trajectory in latent space towards the predicted clean sample $\boldx_0$. Therefore, the reverse process no longer needs to go through all diffusion steps. Instead, inference can be carried out on a sub-sampled schedule,
thus skipping intermediate diffusion steps while preserving the learned marginal distributions. In this work, we retain the stochastic \gls{DDPM} sampler for simplicity  as our aim is to introduce the \gls{JRAA}-\gls{DPS} framework rather than optimizing inference speed, although the same pipeline can be run with a sub-sampled \gls{DDIM} schedule if faster sampling is desired.

In the context of \gls{JRAA}, the \gls{DM} is trained on a collection of paired activity--attenuation images $\boldx = (\boldlambda, \boldmu)$, obtained from the same patient. Since these image pairs share structural similarities, the sampler in \eqref{eq:ddpm-reverse1} iteratively generates a sequence of image pairs $\boldx_t = (\boldlambda_t, \boldmu_t)$ for $t \in {T, T-1, \dots, 0}$, such that the final sample $\boldx_0 = (\boldlambda_0, \boldmu_0)$ also corresponds to the same patient. Consequently, $\boldlambda_0$ and $\boldmu_0$ are expected to exhibit structural consistency (as observed in multi-energy \gls{CT} reconstruction with \gls{DPS} \cite{vazia2024diffusion}). Alternatively, the \glspl{DM} can be trained on each channel independently, which reduces training time but typically yields suboptimal results \cite{phung2024joint}.


\subsection{Diffusion Posterior Sampling}\label{sec:dps}

In presence of a measurement $\boldy$ with known conditional \gls{PDF} $p(\boldy | \boldx)$ given $\boldx$, such as for example with the \gls{PET} model \eqref{eq:poisson}~\&~\eqref{eq:ybar},  it is possible to leverage \glspl{DM} to sample $\boldx = \boldx_0$ from the posterior $p(\boldx | \boldy)$ rather than the prior $p(\boldx)$ \cite{chung2023diffusion}. This can be achieved using  the conditional score given by Bayes' formula,
\begin{equation}\label{eq:bayes}
	\nabla_{\boldx_t} \log p_t (\boldx_t \mid \boldy) = \nabla \log p_t (\boldx_t) + \nabla_{\boldx_t} \log p (\boldy \mid \boldx_t)
\end{equation}
where $p (\boldx_t | \boldy)$ and $p (\boldy | \boldx_t)$ model the dependence between $\boldy$ and $\boldx_t$ through the entire Markov chain $\boldx=\boldx_0,\boldx_1,\dots,\boldx_t$ with \eqref{eq:ddpm-sample}.  Since $\nabla_{\boldx_t} \log p (\boldy | \boldx_t)$ is intractable, the following approximation is used:
\begin{equation}\label{eq:laplace}
 	\nabla_{\boldx_t} \log p (\boldy \mid \boldx_t) \simeq \nabla_{\boldx_t} \log p \big(\boldy \mid \boldx = \boldxhat_{\boldtheta} (\boldx_t,t) \big) \, .
\end{equation}
This approximation requires differentiating through $\boldxhat_{\boldtheta}$, although it can be avoided using simplifications \cite{chung2023decomposed,daras2024survey}. 

In the context of \gls{JRAA}, we propose to implement \gls{DPS} at each time $t\in\{T,T-1, \dots, 1\}$ by first sampling an intermediate estimate $\tilde{\boldx}_{t-1} = \left(\tilde{\boldlambda}_{t-1},\tilde{\boldmu}_{t-1}\right)$ from $\boldx_{t}$ with \eqref{eq:ddpm-reverse1}, then by updating the two images with $\nabla_{\boldlambda_t} \log p_t$ and $\nabla_{\boldmu_t} \log p_t$ separately following the strategy by \citeauthor{song2021solving}~\cite{song2021solving},
\begin{align}
	\boldlambda_{t-1} = {} &  \tilde{\boldlambda}_{t-1} +   \zeta_t \nabla_{\boldlambda_t} \log \, p \big(\boldy \mid \boldx = \boldxhat_{\boldtheta} ({\boldx}_t,t)\big) \label{eq:update_act}  \\
	\boldmu_{t-1} = {} &  \tilde{\boldmu}_{t-1} +   \xi_t \nabla_{\boldmu_t} \log \, p \big(\boldy \mid \boldx = \boldxhat_{\boldtheta} ({\boldx}_t,t)\big)\label{eq:update_attn}
\end{align} 
where $\zeta_t>0$ and $\xi_t>0$ can be tuned separately. 


\subsection{3-D Wavelet Diffusion Models}

The utilization of \glspl{DM} to process \gls{3D} volumes is a challenging task as it requires considerably more computational resources \cite{yu2024pet}.

A common workaround consists in using \gls{2D} \glspl{DM} trained on individual slices combined  with a hand-crafted regularization to promote in-between slice consistency \cite{chung2023solving, singh2024score}. Similarly, several \gls{2D} \glspl{DM} can be leveraged across two or three orthogonal directions \cite{lee2023improving}. One can also feed the \gls{DM} adjacent slices to mitigate axial artifacts \cite{gong2024pet, song2024diffusionblend}. Finally, \gls{LDM} use data compression methods to train diffusion in a lower-dimensional latent space \cite{pinaya2022brain, rout2023solving}.

\Glspl{WDM} are a form of \gls{LDM} that are trained in a wavelet coefficient space \cite{friedrich2024wdm} and can be used to scale \glspl{DM} to  large volumes, and have been used in image reconstruction   \cite{depaepe2025solving,depaepe2025adaptive}. In the \gls{WDM} approach proposed by \citeauthor{friedrich2024wdm}~\cite{friedrich2024wdm}, an orthogonal \gls{DWT}, consisting of combinations of the low-pass filter $\frac{1}{\sqrt{2}}[1,1]$ and the high-pass filter $\frac{1}{\sqrt{2}}[1,-1]$ along each spatial dimension, is utilized. The \gls{DWT} (or encoder) $\boldE \colon \calX \to \calZ \triangleq \R^{8 \times \frac{D}{2} \times \frac{W}{2} \times \frac{H}{2}}$ decomposes an image $\boldx \in \calX$ into an eight-channel wavelet coefficient image $\boldz = \{\boldz_{\mathrm{lll}},\boldz_{\mathrm{llh}},\boldz_{\mathrm{lhl}},\boldz_{\mathrm{lhh}},\boldz_{\mathrm{hll}},\boldz_{\mathrm{hlh}},\boldz_{\mathrm{hhl}},\boldz_{\mathrm{hhh}}\}\in\calZ$ ($\rml$ for low-pass and $\rmh$ for high-pass), where each channel has half the spatial resolution of the original image. Although $\calZ$ and $\calX$ have the same algebraic dimension, the structure of $\calZ$ allows to process smaller volumes for training and inference. 

In the following we apply this framework for \gls{JRAA} with $\boldx = (\boldlambda,\boldmu)$. The latent vector is denoted $\boldz\in\calZ\times\calZ$ and is decomposed as  $\boldz = (\boldz^{\mathrm{act}},\boldz^{\mathrm{attn}})$, with $\boldz^{\mathrm{act}}\in\calZ$ and $\boldz^{\mathrm{attn}}\in\calZ$ being respectively the activity ($\boldlambda$) and attenuation ($\boldmu$) latent representations. 
Denoting $\boldF  \colon (\boldlambda,\boldmu) \mapsto (\boldE(\boldlambda),\boldE(\boldmu)) \in \calZ^2$, our \gls{WDM} is trained by applying $\boldF$  to the \gls{DM}  \eqref{eq:ddpm-sample} to generate a sequence of noisy wavelet coefficient vectors $\boldz_t = (\boldz^{\mathrm{act}}_{t},\boldz^{\mathrm{attn}}_{t})= \boldF(\boldlambda_t,\boldmu_t)$, starting from $\boldz_0 = \boldF(\boldx_0)$, $\boldx_0 = (\boldlambda_0,\boldmu_0) \sim p^{\mathrm{data}}$, as 
\begin{equation}\label{eq:wdm-sample}
	\boldz_t  =  \sqrt{\alpha_t}\boldz_{t-1}  +  \sqrt{1 - \alpha_t}  \boldeta_t  \, ,\quad \boldeta_t \sim \mathcal{N}\left(\boldzero_{\calZ^2} ,  \boldI_{\calZ^2}\right)
\end{equation}
where we used the orthogonality of $\boldF$ so that  $\boldeta_t = \boldF(\boldepsilon_t) \sim \boldepsilon_t $. A denoiser  \gls{CNN} $\boldzhat_{\boldtheta} \colon \calZ\times\calZ \times [0, T] \to \calZ\times\calZ$ similar to   $\boldxhat_{\boldtheta}$ (cf. \eqref{eq:x0nn}) can be trained and used to sample $\boldz_{t-1}$ from $\boldz_{t}$ similarly to \eqref{eq:ddpm-sample}.

Thus, it is possible to utilize the \gls{DPS} scheme described in Section~\ref{sec:dps} to leverage the measurement data using an update scheme similar to \eqref{eq:update_act} and \eqref{eq:update_attn}:
\begin{align}
	\boldz_{t-1}^{\mathrm{act}} = {} &  \tilde{\boldz}_{t-1}^{\mathrm{act}} \nonumber \\
	& \hspace{-7pt} +   \zeta_t \nabla_{\boldz_t^{\mathrm{act}}} \log \, p \Big(\boldy \,\, \big| \,\, \boldx = \boldF^{-1} \left(   \boldzhat_{\boldtheta}   \left({\boldz}_t^{\mathrm{act}}   , {\boldz}_t^{\mathrm{attn}} ,t   \right)          \right)  \Big) \label{eq:update_zact}  \\
	\boldz_{t-1}^{\mathrm{attn}} = {} &  \tilde{\boldz}_{t-1}^{\mathrm{attn}} \nonumber \\
	& \hspace{-7pt} +   \xi_t \nabla_{\boldz_t^{\mathrm{attn}}} \log \, p \Big(  \boldy \,\, \big| \,\, \boldx = \boldF^{-1} \left(   \boldzhat_{\boldtheta}   \left( {\boldz}_t^{\mathrm{act}} , {\boldz}_t^{\mathrm{attn}} ,t  \right)      \right) \Big)   \label{eq:update_zattn} 
\end{align} 
where $\tilde{\boldz}_{t-1} = (\tilde{\boldz}^{\mathrm{act}}_{t-1}, \tilde{\boldz}^{\mathrm{attn}}_{t-1} ) $ is sampled from $\boldz_{t}$ using the \gls{DDPM} scheme described in Section~\ref{sec:dm}.


\subsection{Joint Scatter Estimation}

\Gls{PET} reconstruction requires corrections for scanner sensitivity, scatter, and random coincidences. While sensitivity and random events can be estimated from the scanner and raw measurements \cite{brasse2005correction}, most scatter estimation methods utilize the attenuation map, such as \gls{SSS} \cite{watson2000new} and Monte-Carlo simulations \cite{holdsworth2002performance}. However, in absence of a pre-estimated attenuation map, scatter must be estimated differently. \Gls{AI} approaches have been developed \cite{laurent2023pet,laurent2025evaluation}, but they require \glspl{ACF}. Alternatively,  an energy-based scatter estimation technique could be used \cite{hamill2024energy}. In this work, we propose to deploy a joint scatter estimation method in which scatter is progressively refined during \gls{DPS}, similarly to the approach proposed by \citeauthor{lorenzon2024joint}~\cite{lorenzon2024joint}. 

Each \gls{LOR} $i\in\{ 1,\dots,n_{\rml}  \}$ corresponds to a ring pair $\nu_i$, an angle of view $\varphi_i$ and a radial position $u_i$. 
The scatter vector $\bolds = [s_1,\dots,s_{n_{\rml}}]\transp$ (the \gls{TOF} index was omitted for conciseness) can approximated by a collection of Gaussian-like functions of the radial position $u$, i.e.,
\begin{equation}\label{eq:scatter_model}
	\hat{s}_i (\boldphi) = a_{\nu_i, \varphi_i} \exp \left( -\frac{(u_i - m_{\nu_i, \varphi_i})^2}{2\sigma_{\nu_i, \varphi_i}^2}       \right)
\end{equation}
where $a_{\nu_i, \varphi_i}$ is a scaling factor and $m_{\nu_i, \varphi_i}$ and $\sigma_{\nu_i, \varphi_i}$ are respectively the mean and the \gls{STD}, and the parameters are regrouped in a vector $\boldphi = \left\{ (a_{\nu,\varphi} , m_{\nu, \varphi} , \sigma_{\nu, \varphi} )  , \,  \nu=1,\dots,{n_\rmp} , \varphi = 0, \frac{\pi}{n_\rma} ,\dots, \frac{(n_\rma-1) \pi}{n_\rma} \right\}$, $n_\rmp$ and $n_\rma$ being respectively the number of ring pairs and angles of view.

The expected measurement and conditional \gls{PDF} are then respectively rewritten $\boldybar(\boldx,\boldphi)$ and $p(\boldy \mid \boldx ,\boldphi)$ to reflect the dependency in $\boldphi$ through $\hat{\bolds} (\boldphi) \triangleq [\hat{s}_1 (\boldphi),\dots,\hat{s}_{n_{\rml}} (\boldphi)]\transp $. At each step $t$ of the \gls{DPS} process, we propose to compute $\boldphi_{t-1}$ from $\boldphi_{t}$ using a gradient ascent step of the form
\begin{align}
	\boldphi_{t-1} = {} &\boldphi_{t} +  \gamma_t \nabla_{\boldphi_{t} } \log \, p\left(\boldy \mid \boldx = \boldF^{-1} \left(\boldzhat_{\boldtheta}(\boldz_t, t)\right),\boldphi_t\right) \nonumber \\
	{} &- \delta_t \nabla H (\boldphi_{t} ) \label{eq:scatter_update}
\end{align} 
where $\gamma_t,\delta_t > 0$ and $H$ is a quadratic regularizer that promotes smoothness along the angle $\varphi$ axis.


\subsection{Summary}

Our sampling method, namely \gls{JRAA}-\gls{DPS}, is summarized in Algorithm~\ref{algo:dps}. Step~\ref{step:grad_act} and \ref{step:grad_attn} respectively correspond to \eqref{eq:update_zact} and \eqref{eq:update_zattn} where the gradient is computed by differentiating through $\boldxhat_{0}$ and evaluating at ${\boldz}_{t} $. The scatter parameter update \eqref{eq:scatter_update}  is performed in Step~\ref{step:scatter}. A similar framework was proposed by \citeauthor{chung2023parallel}~\cite{chung2023parallel} for blind inverse problems. 

\begin{algorithm}
	\caption{\gls{3D} \gls{JRAA}-\gls{DPS} sampling}\label{algo:dps}
	\begin{algorithmic}[1]
		\Require $T$, $\bm{y}$, $\{ \alpha_t \}_{t=1}^T$, $\{ \zeta_t \}_{t=1}^T$, $\{ \xi_t \}_{t=1}^T$, $\{ \sigma_t \}_{t=1}^T$, $\{ \gamma_t \}_{t=1}^T$, $\{ \delta_t \}_{t=1}^T$, $\boldphi_{\mathrm{init}}$
		\State $\boldz_T \gets (\boldz_T^{\mathrm{act}}  , \boldz_T^{\mathrm{attn}}    ) \sim \mathcal{N}(\boldzero_{\calZ\times\calZ}, \boldI_{\calZ\times\calZ})$
		\State  $ \boldphi_{T} \gets \boldphi_{\mathrm{init}} $ 
		\For{$t = T $ \textbf{to} $1$}
		    \State $\boldzhat_0 \gets \hat{\boldz}_{\boldtheta}(\boldz_t, t)$
		    \State $\boldeta \sim \mathcal{N}(\boldzero_{\calZ\times\calZ}, \boldI_{\calZ\times\calZ})$
		    \State $\tilde{\boldz}_{t-1} \gets \frac{\sqrt{\alpha_t} (1-\bar{\alpha}_{t-1})}{1-\bar{\alpha}_t} \boldz_t + \frac{\sqrt{\bar{\alpha}_{t-1}  } \beta_t}{1 - \bar{\alpha}_t} \boldzhat_0 + \sigma_t \boldeta$
		    \State $\boldxhat_{0} \gets \boldF^{-1}  (  \hat{\boldz}_{0}  ) $
		    \State $\boldz_{t-1}^\mathrm{act} \gets \tilde{\boldz}_{t-1}^\mathrm{act}  + \zeta_t \nabla_{\boldz_{t}^\mathrm{act}  } \log p \big(\bm{y} \mid \boldx = \boldxhat_{0}(\boldz_t), \boldphi_t\big)  $    \label{step:grad_act}
		    \State $\boldz_{t-1}^\mathrm{attn} \gets \tilde{\boldz}_{t-1}^\mathrm{attn}  + \xi_t \nabla_{\boldz_{t}^\mathrm{attn}  } \log p \big(\bm{y} \mid \boldx = \boldxhat_{0}(\boldz_t), \boldphi_t\big)  $    \label{step:grad_attn}
		    \State $\boldphi_{t-1} \gets \boldphi_{t} +  \gamma_t \nabla_{\boldphi_{t} } \log p\big(\boldy \mid \boldx = \boldxhat_{0}(\boldz_t), \boldphi_t\big) - \delta_t \nabla H (\boldphi_{t} ) $  \label{step:scatter}
		    \State $\boldz_{t-1} \gets (\boldz_{t-1}^\mathrm{act},\boldz_{t-1}^\mathrm{attn})$
		\EndFor
		\State  \Return $\boldF^{-1}(\boldz_0)$
	\end{algorithmic}
\end{algorithm}

To accelerate the reconstruction, step~\ref{step:grad_act} and step~\ref{step:grad_attn} can be implemented with different projection subsets for each $t$, in a similar fashion as with \gls{OSEM} (cf. Section~\ref{sec:experiments}).

The proposed framework can reconstruct data with diverse acquisition characteristics, provided that these are appropriately modeled. In particular, factors such as scan duration (i.e., noise level), scanner geometry, \gls{TOF} information, normalization, and other corrections are explicitly incorporated into the forward model \eqref{eq:ybar}, allowing a single trained \gls{DM} to be applied across different settings without retraining. Furthermore, the joint scatter estimation step is optional and can be omitted when scatter is negligible or when a scatter estimate is available, resulting in a simplified sampling procedure.

	\section{Experiments}\label{sec:experiments}

Model training and reconstruction were implemented with PyTorch. The hyperparameters of \gls{JRAA}-\gls{DPS} were finely tuned to optimize the output images. We used parallelproj \cite{schramm2024parallelproj} to perform projections and backprojections with 4-mm \gls{FWHM} detector resolution. \Gls{JRAA}-\gls{DPS} was implemented with 10 subsets for the activity and attenuation updates (Step~\ref{step:grad_act} and Step~\ref{step:grad_attn} in Algorithm~\ref{algo:dps}).

We evaluated our method on both simulated (\textit{Experiment 1} in Section~\ref{subsec:exp1}) and real \gls{PET} data (\textit{Experiment 2} in Section~\ref{subsec:exp2}).

Retrospective patient data were used for model training, raw data simulation, and image reconstruction. All patients provided informed consent for the use of their data.


\subsection{Patient Data Images}

We used 360 \gls{FDG} \gls{PET}/\gls{CT} patient \gls{3D} images with a 500-mm \gls{FOV} provided by CHU Poitiers to conduct our experiments. \Gls{PET} images were first upsampled to match the 1.50\texttimes{}0.97\texttimes{}0.97-mm\textsuperscript{3} \gls{CT} resolution. \Glsxtrshort{CT} scans were converted from \glsxtrshort{HU} to 511-keV attenuation maps and beds were removed \cite{burger2002pet}. The images were then downsampled to a 2.00\texttimes{}1.95\texttimes{}1.95-mm\textsuperscript{3} resolution (256\texttimes256 slices). Vertically trimming from shoulders to pelvis instead of head to mid-thighs was applied. Finally, \gls{PET} images were clipped to $1\times10^5$ Bq/mL while attenuation maps were clipped to $0.025$ mm\textsuperscript{-1} to remove outliers.  


\subsection{Model Training}

\subsubsection{Normalization}

Attenuation maps were linearly mapped to the $[-1, 1]$ range while \gls{PET} images were normalized as
\begin{equation}
    \boldlambda_{\mathrm{norm}} = 3 \cdot \sqrt[3]{\frac{\boldlambda}{\lambda_{\mathrm{max}}}} - 1 
\end{equation}
where $\lambda_{\mathrm{max}} = 1\times10^5$ and $\boldlambda_{\mathrm{norm}}$ is the normalized \gls{PET} image. While attenuation map values occupy a relatively compact physical range, \gls{PET} images typically have a wide and a right-skewed intensity distribution. The cube-root transform compresses the wide dynamic range of \gls{PET}, limiting the influence of a few high uptake voxels while retaining contrast in lower-activity regions. The resulting normalized \gls{PET} range of [-1, 2] aims to make the activity and attenuation histograms comparable, and to facilitate training on the activity--attenuation image pairs. This normalization strategy is similar to the logarithmic linear normalization proposed by \citeauthor{sun2025ct}~\cite{sun2025ct}, which was shown to be effective for \gls{DM} training on \gls{PET} images by compressing the dynamic range.

\subsubsection{Training parameters}

\begin{table}[ht]
	\centering
	\footnotesize
	\begin{tabular}{llc}
		\toprule
        Method & & \\
        \midrule
		\multirow[c]{4}{*}{JRAA-DPS} & $T$ & $1000$ \\
	    & $\beta_1$ & $10^{-4}$ \\
	    & $\beta_T$ & $0.02$ \\
        & Variance schedule & Linear \\
		\bottomrule
	\end{tabular}
	\caption{Diffusion hyperparameters ($\beta_t = 1 - \alpha_t$).}
	\label{tab:hp-diffusion}
\end{table}

We used 300 patient cases for training and 30 for model validation. We extracted 128\texttimes{}256\texttimes{}256 sub-volumes, resulting in  1,168 volumes for training and 112 for validation. \Gls{JRAA}-\gls{DPS} uses the same diffusion hyperparameters as the original DDPM implementation by \citeauthor{ho2020denoising}~\cite{ho2020denoising} (cf. Table~\ref{tab:hp-diffusion}). Volumes were grouped in batches of 4, then training was performed using the AdamW optimizer with a learning rate of $1 \times 10^{-5}$. We selected the checkpoint with the lowest validation loss as the final model (about 600 epochs). Training data augmentation with random flips in all directions and random affine transformations is performed with TorchIO. Training and reconstructions were carried out on a single 48-GB GPU.


\subsection{Methods for Comparison}

We compared our method with two \gls{JRAA} reconstruction approaches: (i) \gls{MLAA} \cite{rezaei2012simultaneous}, and (ii) the \gls{DL}-based approach proposed by \citeauthor{toyonaga2022deep}~\cite{toyonaga2022deep} which we refer to as \gls{MLAA}-UNet.
The latter uses a U-Net to produce a pseudo-attenuation map from the \gls{MLAA} attenuation estimate, which is then used to perform \gls{AC} reconstruction via \gls{OSEM}.

In this study, we did not consider \gls{JRAA}-\gls{DPS} with \gls{DM} priors trained separately on $\boldlambda$ and $\boldmu$, as prior results have shown that a jointly trained prior over $(\boldlambda, \boldmu)$ consistently achieves better performance \cite{phung2024joint}.

To avoid strong influence of isolated voxels with high values, we clipped each activity reconstruction to the 99.9\% quantile of the reference activity image.

\subsection{Experiment 1---Simulated Data}\label{subsec:exp1}

\subsubsection{Raw Data Simulation and Reconstruction}

We used 30 patients for testing, cropping each volume pair $\boldx = (\boldlambda,\boldmu)$ to  128\texttimes{}256\texttimes{}256 voxels. Emission data $\boldy$ were simulated following \eqref{eq:poisson} from activity--attenuation pairs. 

We varied scan duration $\tau$ to simulate different count levels. \Gls{HC} data were simulated with a scan duration $\tau = 1\times10^{-5}$, and \gls{LC} data were simulated with $\tau = 1\times10^{-7}$ (1\%  of the counts). Random and scatter events were not included in the simulations.

The parallelproj projector used in simulations included 48 rings, each with 300 detectors, and a scanner diameter of 700~mm. The maximum ring difference was set to 44 rings and a radial trim of 30 \glspl{LOR} was applied.
Each projection consists of 241 radial elements, 150  angles of view, and 2,292 sinograms. When applicable, projections also include 13 \gls{TOF} bins with a 58-mm ($\approx$400~ps) \gls{FWHM} \gls{TOF} resolution.

\Gls{MLAA} and \gls{MLAA}-UNet were evaluated with \gls{TOF} data only, while \Gls{JRAA}-\gls{DPS} was evaluated on both \gls{TOF} and non-\gls{TOF} data for reconstruction. For \gls{MLAA} (and subsequently \gls{MLAA}-UNet), attenuation was initialized and updated within a predefined support region, which is obtained from an initial non-\gls{AC} \gls{TOF} \gls{PET} reconstruction. Reconstruction hyperparameters are summarized in Table~\ref{tab:hp-simulations}.

\subsubsection{Quantitative Analysis}\label{sec:quantanal}

We used the \gls{PSNR}, the \gls{SSIM} and the \gls{NRMSE}, computed with respect to the \gls{GT} reference image, using $\mathtt{peak\_signal\_noise\_ratio}$, $\mathtt{structural\_similarity}$ and $\mathtt{normalized\_root\_mse}$ from the Python package scikit-image.

\begin{table}[ht]
	\centering
	\footnotesize
	\begin{tabular}{llcc}
		\toprule
		Method & & HC & LC \\
		\midrule
		MLAA (TOF) & Nb. iterations & 15 & 5 \\
		\thinmidrule
		\multirow[c]{2}{*}{MLAA-UNet (TOF)} & Nb. subsets & 1 & 1 \\
		& Nb. iterations & 15 & 5 \\
        \thinmidrule
		\multirow[c]{3}{*}{JRAA-DPS (TOF)} & Nb. subsets & 10 & 10\\
        & Measurement contrib. $\zeta_t$ & 1 & 0.3 \\
		& Measurement contrib. $\xi_t$ & 0.6 & 0.2\\
        \thinmidrule
		\multirow[c]{3}{*}{JRAA-DPS (non-TOF)} & Nb. subsets & 10 & 10\\
        & Measurement contrib. $\zeta_t$ & 0.6 & 0.3 \\
		& Measurement contrib. $\xi_t$ & 0.4 & 0.2 \\
		\bottomrule
	\end{tabular}
	\caption{\textit{Experiment 1}---Reconstruction hyperparameters.}
	\label{tab:hp-simulations}
\end{table}

We quantified the bias-variance trade-off of \gls{MLAA}, \gls{MLAA}-UNet and \gls{JRAA}-\gls{DPS} using \gls{TOF} measurements for different values of the hyperparameters. We also included results for \gls{OSEM} \gls{TOF} reconstruction with the true attenuation map. For \gls{MLAA}, \gls{MLAA}-UNet and \gls{OSEM} the number of iterations was swept from 1 to 30 (increments of 1 between 1 and 5, then increments of 5). All three methods used a single subset in these experiments. For \gls{JRAA}-\gls{DPS} we swept the measurement-contribution parameter $\zeta_t$ from 0.5 to 5 while holding $\xi_t=0.5$ when assessing activity reconstructions, and swept $\xi_t$ from 0.5 to 5 while holding $\zeta_t=0.5$ for attenuation reconstructions. Normalized absolute bias and normalized \gls{STD}, both averaged over the entire image, were computed over 30 independent Poisson noise realizations of the same \gls{GT} reference activity and attenuation images for each configuration.

To evaluate lesion recovery, we inserted a synthetic 15-mm-diameter hot lesion in the liver of a single patient's dataset. We use two complementary metrics: the \gls{RC} measures how closely the reconstructed lesion intensity matches the ground-truth lesion intensity, while the \gls{CRC} quantifies lesion--background contrast recovery. For each method---\gls{OSEM} (with the true patient attenuation map), \gls{MLAA}, \gls{MLAA}-UNet and \gls{JRAA}-\gls{DPS}---\gls{RC} and \gls{CRC} were averaged over the 30 noise realizations. \Gls{OSEM}, \gls{MLAA} and \gls{MLAA}-UNet used a single subset and iterations ranging from 1 to 30. In \gls{JRAA}-\gls{DPS}, $\zeta_t$ varied from 0.5 to 5 with $\xi_t=0.5$.

As \gls{DDPM} is a stochastic process, the reconstruction is non-deterministic, and a single measurement can therefore yield multiple reconstructions. To assess the variability of the method, we performed 30 reconstructions for each count setting from a single measurement $\boldy$, both with and without \gls{TOF} information. We then computed the mean reconstruction and the \gls{STD} maps based of these 30 reconstructions.


\subsection{Experiment 2---Clinical Data}\label{subsec:exp2}

To demonstrate the clinical applicability of \gls{JRAA}-\gls{DPS}, we reconstructed non-\gls{TOF} \gls{FDG} \gls{PET} data acquired with a Siemens Biograph mMR. The scanner was modeled using parallelproj and corrections such as normalization and random events were computed using the system's tools. The reconstructed volumes are 128\texttimes{}256\texttimes{}256 voxels with voxel dimensions of 2.03\texttimes{}2.09\texttimes{}2.09-mm\textsuperscript{3}. 

A reference \gls{PET} image was reconstructed using \gls{OSEM} (3 iterations, 10 subsets), applying the \gls{MR}-derived \glspl{ACF} for attenuation correction. 

\Gls{MLAA} and \gls{MLAA}-UNet (as well as the \gls{OSEM} reference image) used the system's build-in scatter estimate (attenuation-based)
to make comparisons as favorable as possible. \Gls{JRAA}-\gls{DPS} was implemented with the proposed joint scatter estimation strategy. Thus, the reconstruction does not rely on the attenuation-derived scatter estimates.

\Gls{MLAA} attenuation was initialized and updated within a support derived from the reference \gls{MR}-derived attenuation map, as the support of the initial non-\gls{AC} non-\gls{TOF} estimate was too inaccurate. This is a strong prior information that favors \gls{MLAA} and \gls{MLAA}-UNet for this experiment.

We trained \gls{MLAA}-UNet on \gls{MLAA} reconstructions of non-\gls{TOF} raw data simulated with the Biograph mMR geometry. For each training case, \gls{MLAA} used an attenuation support derived from the corresponding reference \gls{CT}-based attenuation map.

Reconstruction hyperparameters are summarized in Table~\ref{tab:hp-realdata}.

The \gls{JRAA}-\gls{DPS} reconstruction was further refined with a single \gls{OSEM} iteration (10 subsets) using the estimated attenuation.

\begin{table}[ht]
    \centering
    \footnotesize
    \begin{tabular}{llc}
        \toprule
        Method & & mMR data  \\
        \midrule
        MLAA (non-TOF) & Nb. iterations & 25 \\
        \thinmidrule
        \multirow[c]{2}{*}{MLAA-UNet (non-TOF)} & Nb. subsets & 10 \\
         & Nb. iterations & 30 \\
        \thinmidrule
		\multirow[c]{3}{*}{JRAA-DPS (non-TOF)} & Nb. subsets & 10\\
        & Measurement contrib. $\zeta_t$ & 1 \\
		& Measurement contrib. $\xi_t$ & 0.6\\
        \bottomrule
    \end{tabular}
    \caption{\textit{Experiment 2}---Reconstruction hyperparameters.}
    \label{tab:hp-realdata}
\end{table}

The bed, present in the \gls{FOV}, produces non-negligible attenuation. Since our model was trained on bedless simulated data, it is not capable of representing bed attenuation. The bed is not visible in \gls{MR} scans, but its attenuation map can be obtained from a transmission scan \cite{bowen2016transmission, farag2021improved}. Denoting $\boldb \in \calX$ the attenuation map of the bed, we modified the \gls{ACF} definition \eqref{eq:attn} as 
\begin{equation}\label{eq:attn_bed}
	a_i(\boldmu)  = \rme^{- [\boldR [ \boldmu + \boldb] ]_i } 
\end{equation}
and incorporated this \gls{ACF} in the forward model \eqref{eq:ybar}. Training on bedless simulations and using an estimated bed attenuation map at inference obviates the need for scanner-specific retraining, since bed designs can vary across systems.
	\section{Evaluation}\label{sec:results}


\subsection{Experiment 1---Simulated Data Reconstructions}\label{sec:eval_sim}

\input{./figures/fig_recons_hc}
\input{./figures/fig_recons_lc1}
\newcommand{\conf}[1]{{\scriptsize $\pm$ #1}}
\begin{table*}[htbp]
	\footnotesize
	\centering
	\begin{tabular}{llllllll}
		\toprule
		& & \multicolumn{3}{c}{Activity image $\boldlambda$} & \multicolumn{3}{c}{Attenuation map $\boldmu$} \\ 
		\cmidrule(lr){3-5} \cmidrule(lr){6-8}
		& Method & PSNR $\uparrow$ & SSIM $\uparrow$ & NRMSE $\downarrow$ & PSNR $\uparrow$ & SSIM $\uparrow$ & NRMSE $\downarrow$ \\
		\midrule
		\multirow[c]{3}{*}{HC} & MLAA (TOF) & 23.43 \conf{1.67} & 0.70 \conf{0.03} & 0.61 \conf{0.06} & 20.07 \conf{0.36} & 0.49 \conf{0.02} & 0.38 \conf{0.02} \\
		& MLAA-UNet (TOF) & 29.23 \conf{1.67} & 0.83 \conf{0.02} & 0.31 \conf{0.03} & \ul{28.22} \conf{0.33} & \ul{0.83} \conf{0.01} & \ul{0.15} \conf{0.01} \\
		& \gls{JRAA}-\gls{DPS} (TOF) & \ul{32.33} \conf{1.36} & \ul{0.88} \conf{0.02} & \ul{0.22} \conf{0.02} & 27.40 \conf{0.31} & 0.82 \conf{0.01} & 0.16 \conf{0.01} \\
		& \gls{JRAA}-\gls{DPS} (non-TOF) & 29.20 \conf{1.43} & 0.83 \conf{0.02} & 0.29 \conf{0.02} & 25.10 \conf{0.47} & 0.77 \conf{0.01} & 0.21 \conf{0.02} \\
		\midrule
		\multirow[c]{4}{*}{LC} & MLAA (TOF) & 17.80 \conf{1.37} & 0.41 \conf{0.03} & 1.43 \conf{0.14} & 17.71 \conf{0.27} & 0.40 \conf{0.01} & 0.50 \conf{0.02} \\
		& MLAA-UNet (TOF) & 20.59 \conf{1.29} & 0.50 \conf{0.04} & 1.04 \conf{0.09} &  \ul{23.84} \conf{0.29} & \ul{0.74} \conf{0.01} & \ul{0.25} \conf{0.01} \\
		& \gls{JRAA}-\gls{DPS} (TOF) & \ul{27.02} \conf{1.29} & \ul{0.77} \conf{0.03} & \ul{0.39} \conf{0.03} & 22.62 \conf{0.29} & 0.70 \conf{0.01} & 0.28 \conf{0.01} \\
		& \gls{JRAA}-\gls{DPS} (non-TOF) & 25.90 \conf{1.39} & 0.75 \conf{0.03} & 0.44 \conf{0.03} & 22.22 \conf{0.42} & 0.69 \conf{0.01} & 0.30 \conf{0.02} \\
		\bottomrule
	\end{tabular}
	\caption{\textit{Experiment 1}---Reconstruction metrics means and 95\% confidence intervals for \gls{HC} and \gls{LC} data.}
	\label{tab:metrics-full}
\end{table*}
\begin{table}[htbp]
	\footnotesize
	\centering
	\begin{tabular}{llr}
		\toprule
		& Method & \makecell[l]{Reconstruction \\ duration (min)}  \\
		\midrule
		\multirow[c]{4}{*}{HC} & MLAA (TOF) & 2.5 \\
		& MLAA-UNet (TOF) & 3.1 \\
		& JRAA-DPS (TOF) & 43.0 \\
		& JRAA-DPS (non-TOF) & 33.0 \\
		\midrule
		\multirow[c]{4}{*}{LC} & MLAA (TOF) & 0.9 \\
		& MLAA-UNet (TOF) & 1.1 \\
		& JRAA-DPS (TOF) & 43.0 \\
		& JRAA-DPS (non-TOF) & 33.0 \\
		\bottomrule
	\end{tabular}
	\caption{
		\textit{Experiment 1}---Average reconstruction durations across test cases. 
	}
	\label{tab:recon-time-test}
\end{table}

\begin{figure}
	\centering
	\begin{tikzpicture}
	\begin{groupplot}[
		group style={
			group size=2 by 2,
			vertical sep=2mm,
			horizontal sep=2mm,
			xlabels at=edge bottom,
			ylabels at=edge left,
		},
		width=0.58\linewidth,
		height=0.50\linewidth,
		xmin=11, xmax=41,
		ymin=0, ymax=1,
		tick align=outside,
		xtick pos=bottom,
		ytick pos=left,
		tick label style={font=\tiny},
		tick style={major tick length=2pt},
		xticklabel style={yshift=2pt},
		yticklabel style={xshift=2pt},
		xlabel style={yshift=3pt},
		ylabel style={yshift=-3pt},
		every axis label/.append style={font=\scriptsize},
		title style={
			font=\scriptsize,
			yshift=-6pt
		},
		every axis plot/.append style={
			thick,
			line cap=round,
			line join=round,
		},
		]
		
		\nextgroupplot[
		xticklabels={},
		ylabel={SSIM (Activity)},
		title={\gls{HC}},
		legend pos=south east,
		legend cell align=left,
		legend style={
			font=\tiny,
			row sep=0pt,
			column sep=3pt,
			inner sep=2pt,
			nodes={inner sep=1pt}
		},
		]
		\addplot[mlaa_points] table[x=mlaa_psnr,y=mlaa_ssim,col sep=space]{figures/psnr_ssim/pet_hc_metrics.dat};
		\addlegendentry{MLAA (TOF)}
		
		\addplot[unet_points] table[x=unet_psnr,y=unet_ssim,col sep=space]{figures/psnr_ssim/pet_hc_metrics.dat};
		\addlegendentry{MLAA-UNet (TOF)}
		
		\addplot[dpsnotof_points, forget plot] table[x=jraadpsnotof_psnr,y=jraadpsnotof_ssim,col sep=space]{figures/psnr_ssim/pet_hc_metrics.dat};
		
		\addplot[dpstof_points, forget plot] table[x=jraadpstof_psnr,y=jraadpstof_ssim,col sep=space]{figures/psnr_ssim/pet_hc_metrics.dat};
		
		\addlegendimage{dpstof_points}
		\addlegendentry{JRAA-DPS (TOF)}
		
		\addlegendimage{dpsnotof_points}
		\addlegendentry{JRAA-DPS (non-TOF)}
		
		\nextgroupplot[
		xticklabels={},
		yticklabels={},
		title={\gls{LC}},
		]
		\addplot[mlaa_points] table[x=mlaa_psnr,y=mlaa_ssim,col sep=space]{figures/psnr_ssim/pet_lc1_metrics.dat};
		\addplot[unet_points] table[x=unet_psnr,y=unet_ssim,col sep=space]{figures/psnr_ssim/pet_lc1_metrics.dat};
		\addplot[dpsnotof_points] table[x=jraadpsnotof_psnr,y=jraadpsnotof_ssim,col sep=space]{figures/psnr_ssim/pet_lc1_metrics.dat};
		\addplot[dpstof_points] table[x=jraadpstof_psnr,y=jraadpstof_ssim,col sep=space]{figures/psnr_ssim/pet_lc1_metrics.dat};
		
		\nextgroupplot[
		ylabel={SSIM (Attenuation)},
		xlabel={PSNR},
		]
		\addplot[mlaa_points] table[x=mlaa_psnr,y=mlaa_ssim,col sep=space]{figures/psnr_ssim/mu_hc_metrics.dat};
		\addplot[unet_points] table[x=unet_psnr,y=unet_ssim,col sep=space]{figures/psnr_ssim/mu_hc_metrics.dat};
		\addplot[dpsnotof_points] table[x=jraadpsnotof_psnr,y=jraadpsnotof_ssim,col sep=space]{figures/psnr_ssim/mu_hc_metrics.dat};
		\addplot[dpstof_points] table[x=jraadpstof_psnr,y=jraadpstof_ssim,col sep=space]{figures/psnr_ssim/mu_hc_metrics.dat};
		
		\nextgroupplot[
		xlabel={PSNR},
		yticklabels={},
		]
		\addplot[mlaa_points] table[x=mlaa_psnr,y=mlaa_ssim,col sep=space]{figures/psnr_ssim/mu_lc1_metrics.dat};
		\addplot[unet_points] table[x=unet_psnr,y=unet_ssim,col sep=space]{figures/psnr_ssim/mu_lc1_metrics.dat};
		\addplot[dpsnotof_points] table[x=jraadpsnotof_psnr,y=jraadpsnotof_ssim,col sep=space]{figures/psnr_ssim/mu_lc1_metrics.dat};
		\addplot[dpstof_points] table[x=jraadpstof_psnr,y=jraadpstof_ssim,col sep=space]{figures/psnr_ssim/mu_lc1_metrics.dat};
		
	\end{groupplot}
\end{tikzpicture}
	\caption{\textit{Experiment 1}---\Gls{SSIM} against \gls{PSNR} plots for 30 test reconstructions.}
	\label{fig:psnr_ssim}
\end{figure}

\begin{figure}
	\centering
	\begin{tikzpicture}
	\begin{groupplot}[
		group style={
			group size=2 by 2,
			vertical sep=3.5mm,
			horizontal sep=2mm,
			xlabels at=edge bottom,
			ylabels at=edge left,
		},
		width=0.58\linewidth,
		height=0.50\linewidth,
		xmin=0,
		ymin=0,
		tick align=outside,
		xtick pos=bottom,
		ytick pos=left,
		tick label style={
			font=\tiny,
			/pgf/number format/fixed,
			/pgf/number format/precision=4,
			/pgf/number format/1000 sep={},
		},
		tick style={major tick length=2pt},
		xticklabel style={yshift=2pt}, 
		yticklabel style={xshift=2pt},
		xlabel style={yshift=3pt},
		ylabel style={yshift=-3pt},
		every axis label/.append style={font=\scriptsize},
		title style={
			font=\scriptsize,
			yshift=-6pt
		},
		every axis plot/.append style={
			thick,
			line cap=round,
			line join=round,
		},
		]
		
		\nextgroupplot[
		ylabel={Absolute bias (Activity)},
		title={\gls{HC}},
		legend pos=south east,
		legend cell align=left,
		legend style={
			font=\tiny,
			row sep=0pt,
			inner sep=2pt,
			nodes={inner sep=1pt}
		},
		ymax=0.6,
		xmax=0.65
		]
		
		\addplot[ref] table[x=variances_pet_mlem,y=biases_pet_mlem,col sep=space]{figures/bias_variance/bias_variance_iterations_tau1e-05_activity.dat};
		\addplot[mlaa] table[x=variances_pet_mlaa,y=biases_pet_mlaa,col sep=space]{figures/bias_variance/bias_variance_iterations_tau1e-05_activity.dat};
		\addplot[unet] table[x=variances_pet_unet,y=biases_pet_unet,col sep=space]{figures/bias_variance/bias_variance_iterations_tau1e-05_activity.dat};
		\addplot[dpstof] table[x=variances_pet_dps,y=biases_pet_dps,col sep=space]{figures/bias_variance/bias_variance_dps_tau1e-05_activity.dat};
		
		\nextgroupplot[
		title={\gls{LC}},
		yticklabels={},
		ymax=0.6,
		]
		
		\addplot[ref] table[x=variances_pet_mlem,y=biases_pet_mlem,col sep=space]{figures/bias_variance/bias_variance_iterations_tau1e-07_activity.dat};
		\addplot[mlaa] table[x=variances_pet_mlaa,y=biases_pet_mlaa,col sep=space]{figures/bias_variance/bias_variance_iterations_tau1e-07_activity.dat};
		\addplot[unet] table[x=variances_pet_unet,y=biases_pet_unet,col sep=space]{figures/bias_variance/bias_variance_iterations_tau1e-07_activity.dat};
		\addplot[dpstof] table[x=variances_pet_dps,y=biases_pet_dps,col sep=space]{figures/bias_variance/bias_variance_dps_tau1e-07_activity.dat};
		
		\nextgroupplot[
		ylabel={Absolute bias (Attenuation)},
		xlabel={\gls{STD}},
		legend pos=north east,
		legend cell align=left,
		legend style={
			font=\tiny,
			row sep=0pt,
			inner sep=2pt,
			nodes={inner sep=1pt}
		},
		legend image post style={
			line cap=round,
			line join=round,
			xscale=0.7
		},
		ymax=0.6,
		]
		\addlegendimage{ref}
		\addlegendentry{OSEM (true $\boldmu$)}
		\addplot[mlaa] table[x=variances_mu_mlaa,y=biases_mu_mlaa,col sep=space]{figures/bias_variance/bias_variance_iterations_tau1e-05_attenuation.dat};
		\addlegendentry{MLAA}
		\addplot[unet] table[x=variances_mu_unet,y=biases_mu_unet,col sep=space]{figures/bias_variance/bias_variance_iterations_tau1e-05_attenuation.dat};
		\addlegendentry{MLAA-UNet}
		\addplot[dpstof] table[x=variances_mu_dps,y=biases_mu_dps,col sep=space]{figures/bias_variance/bias_variance_dps_tau1e-05_attenuation.dat};
		\addlegendentry{JRAA-DPS}
		
		\nextgroupplot[
		xlabel={\gls{STD}},
		yticklabels={},
		ymax=0.6,
		]
		\addplot[mlaa] table[x=variances_mu_mlaa,y=biases_mu_mlaa,col sep=space]{figures/bias_variance/bias_variance_iterations_tau1e-07_attenuation.dat};
		\addplot[unet] table[x=variances_mu_unet,y=biases_mu_unet,col sep=space]{figures/bias_variance/bias_variance_iterations_tau1e-07_attenuation.dat};
		\addplot[dpstof] table[x=variances_mu_dps,y=biases_mu_dps,col sep=space]{figures/bias_variance/bias_variance_dps_tau1e-07_attenuation.dat};
		
	\end{groupplot}
\end{tikzpicture}
	\caption{\textit{Experiment 1}---Normalized absolute bias against normalized \gls{STD}  for different values of the hyperparameters (cf. Section~\ref{sec:quantanal}). All reconstructions used \gls{TOF} data.}
	\label{fig:bias_variance}
\end{figure}

\begin{figure}
	\centering
	\input{figures/fig_lesion_recovery}
	\vspace{-2mm}
	\caption{\textit{Experiment 1}---\Gls{CRC} against \gls{RC} for a synthetic 15-mm-diameter hot lesion added into the liver for different values of the hyperparameters (cf. Section~\ref{sec:quantanal}). The background was defined as a 70-mm-diameter spherical \gls{ROI} surrounding the lesion. Mean reconstructions with \gls{JRAA}-\gls{DPS} are displayed against the \gls{GT} reference. All reconstructions used \gls{TOF} data.}
	\label{fig:lesion_recovery}
\end{figure}

\begin{figure*}[htbp]
	\centering
	\setlength{\tabcolsep}{0pt}
	\renewcommand{\arraystretch}{1}
	
	\begin{tabular}{rccc:cccl}

		& & & & & & & \\[-1.5ex]

		\rotatebox{90}{\parbox[c][0.3\tempdimhb][c]{\tempdimhb}{\centering \scriptsize Mean $\boldlambda$}}
		&
		\includegraphics[width=\tempdimwb,height=\tempdimhb]{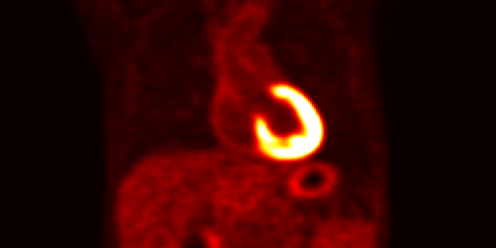}
		&
		\includegraphics[width=\tempdimwb,height=\tempdimhb]{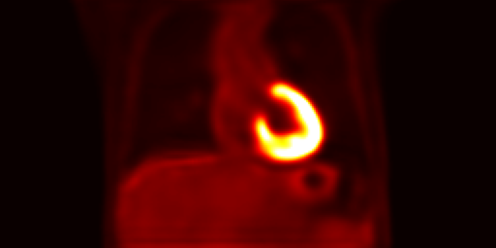}
		&
		\includegraphics[width=\tempdimwb,height=\tempdimhb]{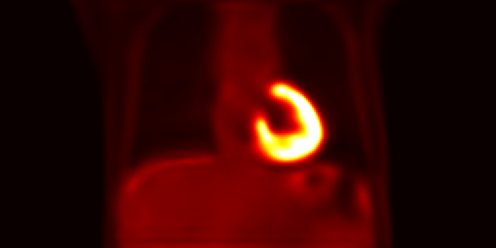}
		&
		\includegraphics[width=\tempdimwb,height=\tempdimhb]{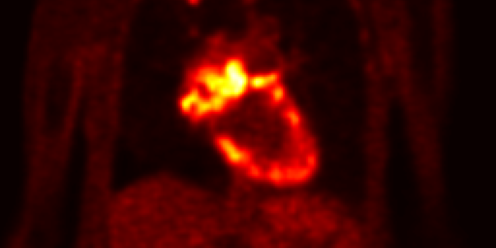}
		&
		\includegraphics[width=\tempdimwb,height=\tempdimhb]{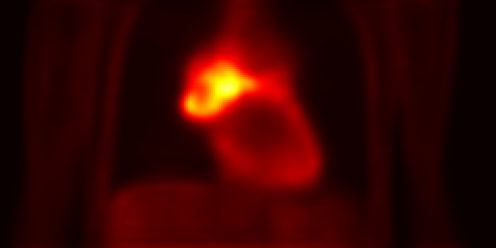}
		&
		\includegraphics[width=\tempdimwb,height=\tempdimhb]{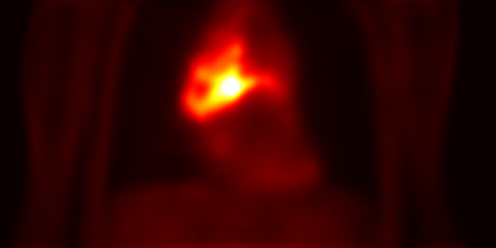}
		&
		\includegraphics[height=0.98\tempdimhb]{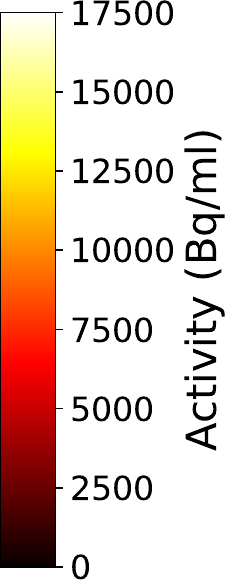}
		\\
		

		\rotatebox{90}{\parbox[c][0.3\tempdimhb][c]{\tempdimhb}{\centering \scriptsize Standard deviation $\boldlambda$}}
		&
		&
		\includegraphics[width=\tempdimwb,height=\tempdimhb]{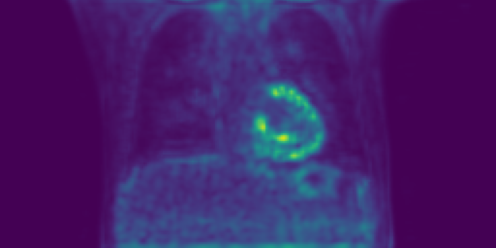}
		&
		\includegraphics[width=\tempdimwb,height=\tempdimhb]{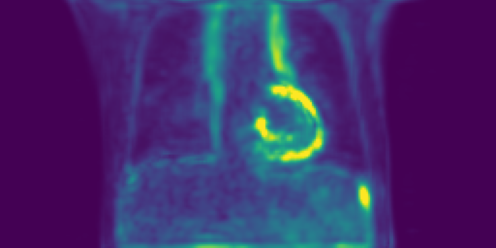}
		&
		&
		\includegraphics[width=\tempdimwb,height=\tempdimhb]{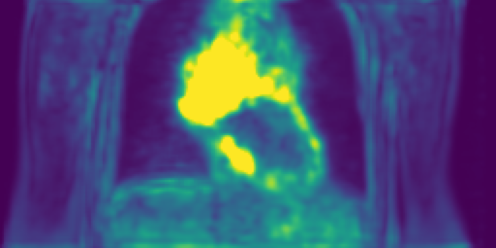}
		&
		\includegraphics[width=\tempdimwb,height=\tempdimhb]{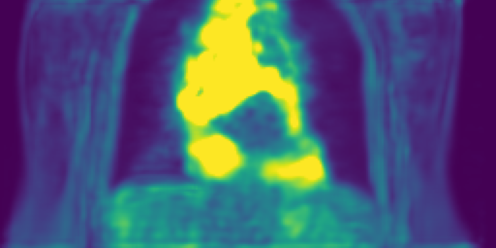}
		&
		\includegraphics[height=0.98\tempdimhb]{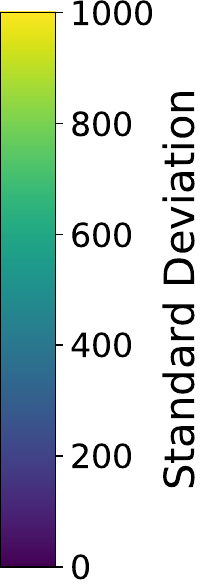}
		\\

		\rotatebox{90}{\parbox[c][0.3\tempdimhb][c]{\tempdimhb}{\centering \scriptsize Mean $\boldmu$}}
		&
		\includegraphics[width=\tempdimwb,height=\tempdimhb]{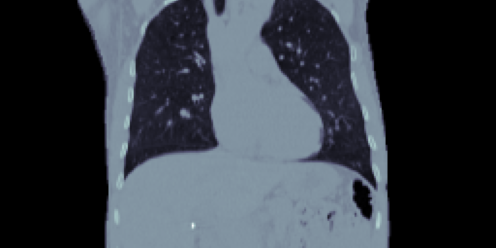}
		&
		\includegraphics[width=\tempdimwb,height=\tempdimhb]{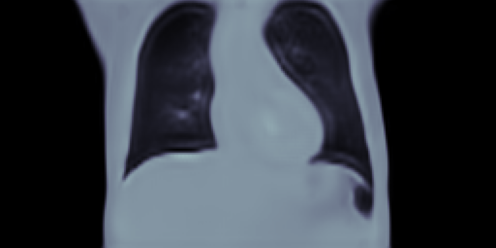}
		&
		\includegraphics[width=\tempdimwb,height=\tempdimhb]{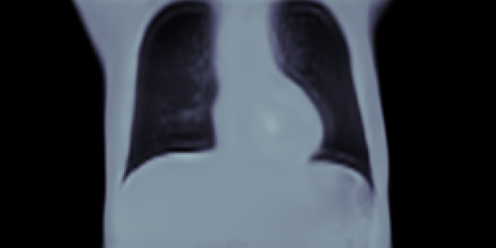}
		&
		\includegraphics[width=\tempdimwb,height=\tempdimhb]{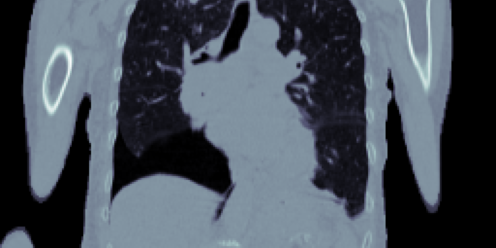}
		&
		\includegraphics[width=\tempdimwb,height=\tempdimhb]{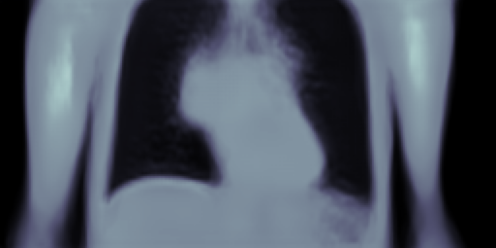}
		&
		\includegraphics[width=\tempdimwb,height=\tempdimhb]{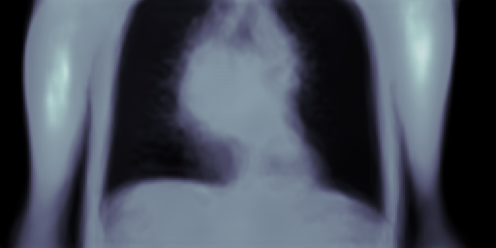}
		&
		\includegraphics[height=0.98\tempdimhb]{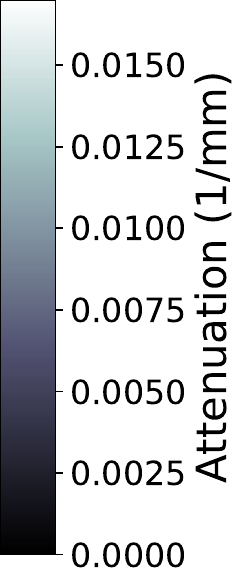}
		\\
		

		\rotatebox{90}{\parbox[c][0.3\tempdimhb][c]{\tempdimhb}{\centering \scriptsize Standard deviation $\boldmu$}}
		&
		\subfloat[HC reference \label{subfig:onetomany_hc}]{%
			\begin{tikzpicture}
				\node[inner sep=0pt, opacity=0] {
					{\rule{\tempdimwb}{\tempdimhb}}
				};
			\end{tikzpicture}
		}
		&
		\subfloat[HC (TOF) \label{subfig:onetomany_hc-tof}]{%
			\includegraphics[width=\tempdimwb,height=\tempdimhb]{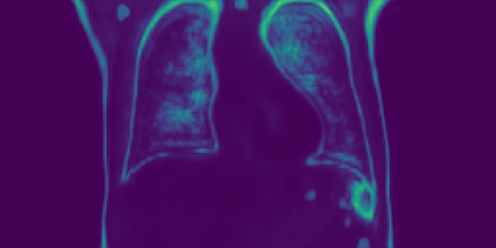}
		}
		&
		\subfloat[HC (non-TOF) \label{subfig:onetomany_hc-notof}]{%
			\includegraphics[width=\tempdimwb,height=\tempdimhb]{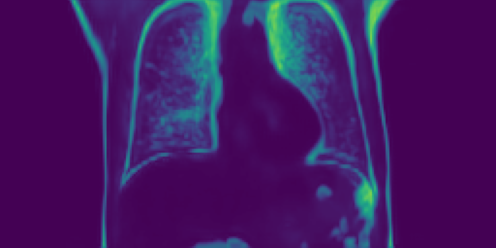}
		}
		&
		\subfloat[LC reference \label{subfig:onetomany_lc}]{%
			\begin{tikzpicture}
				\node[inner sep=0pt, opacity=0] {
					{\rule{\tempdimwb}{\tempdimhb}}
				};
			\end{tikzpicture}
		}
		&
		\subfloat[LC (TOF) \label{subfig:onetomany_lc-tof}]{%
			\includegraphics[width=\tempdimwb,height=\tempdimhb]{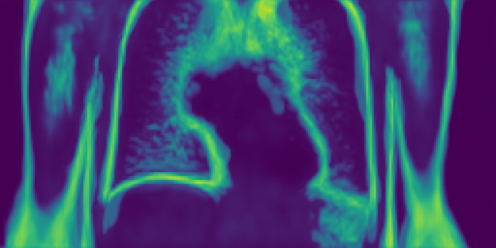}
		}
		&
		\subfloat[LC (non-TOF) \label{subfig:onetomany_lc-notof}]{%
			\includegraphics[width=\tempdimwb,height=\tempdimhb]{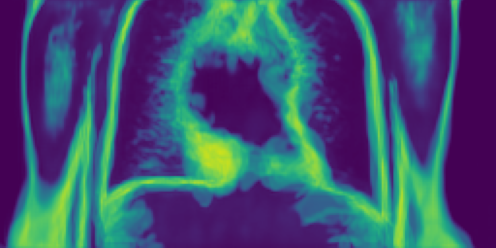}
		}
		&
		\includegraphics[height=0.98\tempdimhb]{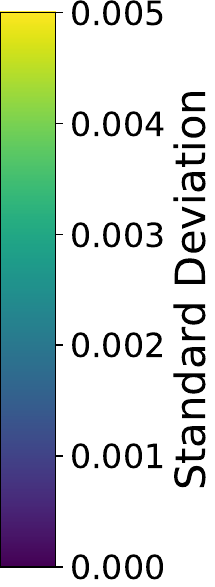}
	\end{tabular}
	
	\caption{\textit{Experiment 1}---Mean and standard deviation maps of 30 JRAA-DPS reconstructions of a single measurement.}
	\label{fig:onetomany}
\end{figure*}

\Gls{HC} and \gls{LC} reconstructions are respectively shown in Figures~\ref{fig:hc_imgs} and \ref{fig:lc1_imgs} for two different patients. Metrics averaged over the testing dataset are shown in Table~\ref{tab:metrics-full}, while scatter plots of \gls{SSIM} against \gls{PSNR} are shown in Figure~\ref{fig:psnr_ssim}.

With regards to \gls{PSNR}, \gls{SSIM} and \gls{NRMSE}, \gls{JRAA}-\gls{DPS} outperforms other methods for activity reconstruction in both count settings. Omitting \gls{TOF} information degrades \gls{JRAA}-\gls{DPS} performance, but it still achieves better reconstructions than \gls{MLAA} and \gls{MLAA}-UNet with \gls{TOF}. 

In the \gls{LC} setting, \gls{MLAA}-UNet outperforms \gls{JRAA}-\gls{DPS} on attenuation map reconstruction across the full test set (cf. Table~\ref{tab:metrics-full}). \Gls{MLAA}-UNet reconstructs a smooth attenuation map while \gls{JRAA}-\gls{DPS} adds in some details which can be erroneous, thus driving metrics down (cf. Figures~\ref{subfig:lc1_imgs-unet} and \ref{subfig:lc1_imgs-ours_tof}). In spite of that, \gls{JRAA}-\gls{DPS} recovers the activity image better (cf. Table~\ref{tab:metrics-full} and Figure~\ref{subfig:lc1_imgs-ours_tof}), suggesting those erroneous details do not alter significantly the quality of the reconstructed activity image.

\Gls{MLAA} and \gls{MLAA}-UNet activity reconstructions suffer from noise amplification and \gls{PVE} as a result of the limited resolution of the imaging system. This is especially noticeable in the \gls{LC} setting (cf. Figure~\ref{subfig:lc1_imgs-mlaa} and \ref{subfig:lc1_imgs-unet}). Conversely, \gls{JRAA}-\gls{DPS} with \gls{TOF} recovers smooth and noiseless activity images (cf. Figure~\ref{subfig:lc1_imgs-ours_tof}). 

\Gls{JRAA}-\gls{DPS} without \gls{TOF} in the \gls{HC} setting still yields a reasonably accurate activity reconstruction (cf. Figure~\ref{subfig:hc_imgs-ours_notof} and Figure~\ref{fig:psnr_ssim}). In the \gls{LC} setting, the image is degraded and the reconstructed attenuation map is visibly distorted, highlighting the limited applicability of \gls{JRAA}-\gls{DPS} in such conditions (cf. Figure~\ref{subfig:lc1_imgs-ours_notof}).

Absolute bias is shown against \gls{STD} in Figure~\ref{fig:bias_variance}, for the \gls{HC} and \gls{LC} settings. \Gls{JRAA}-\gls{DPS} achieves a better bias--variance trade-off than \gls{MLAA} and \gls{MLAA}-UNet in both count settings for activity reconstruction, although the performance gain is clearer in the \gls{LC} setting. \gls{JRAA}-\gls{DPS} has lower bias than \gls{MLAA} for attenuation map reconstruction, but its variance is on the higher end. However, \gls{STD} of \gls{JRAA}-\gls{DPS} attenuation reconstructions remains similar to the \gls{STD} of \gls{JRAA}-\gls{DPS} activity reconstructions.

The \gls{RC} and the \gls{CRC} of the added hot lesion is shown in Figure~\ref{fig:lesion_recovery}. While \gls{MLAA} achieves the lowest \gls{RC} and \gls{CRC} in the \gls{HC} setting, \gls{JRAA}-\gls{DPS} is able to accurately recover both the intensity and the contrast in the studied \gls{ROI}. Furthermore, our proposed method seems robust to hyperparameter selection in \gls{HC}, as varying $\zeta_t$ produces a limited spread in \gls{RC} and \gls{CRC} values. In an \gls{LC} setting, \gls{JRAA}-\gls{DPS} outperforms both \gls{MLAA} and \gls{MLAA}-UNet, provided the measurement-contribution parameter $\zeta_t$ is sufficiently high.

Reconstruction variability of \gls{JRAA}-\gls{DPS} is illustrated in Figure~\ref{fig:onetomany}, where mean and \gls{STD} images are computed from 30 reconstructions obtained from a single measurement under both count settings. We observe that variability increases in the \gls{LC} regime and in the absence of \gls{TOF} information. This behavior arises from the greater ill-posedness of the joint reconstruction problem under these acquisition settings. In the \gls{LC} regime, the measurements provide weaker statistical support, while non-TOF data provides non-localized information along each \gls{LOR}. As a result, a broader set of activity and attenuation distributions can be consistent with the measured data.

In these more ill-posed settings, measurement contributions $\zeta_t$ and $\xi_t$ are reduced to limit noise in the reconstructions (cf. Table~\ref{tab:hp-simulations}). Therefore, the data-consistency terms provide weaker guidance, and the relative influence of the DM-based stochastic prior increases. This increased reliance on the stochastic prior leads to a broader set of plausible reconstructions and this higher sample-to-sample variability. We further observe that variability is highest in regions of elevated tracer uptake in the reconstructed \gls{PET} images, whereas it is most pronounced at body--background and lung--body boundaries in the attenuation maps. Practically, this suggests \gls{LC} and non-\gls{TOF} reconstruction should be interpreted with greater caution in these regions, and the \gls{STD} images provide useful indicators of regions where the reconstruction is less constrained by the data.

Finally, sampling with \gls{DM} is known to be computationally expensive and we observe the proposed \gls{JRAA}-\gls{DPS} method is approximately 14 times slower than \gls{MLAA} and \gls{MLAA}-UNet in the \gls{HC} setting with \gls{TOF} (cf. Table~\ref{tab:recon-time-test}).


\subsection{Experiment 2---Clinical Data Reconstructions}\label{sec:eval_patient}

\input{./figures/fig_recons_realdata}

\begin{figure}
	\centering
    \begin{tikzpicture}
    \begin{groupplot}[
        group style={
            group size=1 by 2,
            vertical sep=2mm,      
            xlabels at=edge bottom 
        },
        width=0.98\linewidth,
        height=0.43\linewidth,
        xmin=0, xmax=256,
        ymin=0,
        xtick={0,50,...,256},
        tick align=outside,
        xtick pos=bottom,
        ytick pos=left,
        tick label style={
            font=\tiny,
            /pgf/number format/fixed,
            /pgf/number format/precision=4,
            /pgf/number format/1000 sep={},
        },
        tick style={major tick length=2pt},
        xticklabel style={yshift=2pt}, 
        yticklabel style={xshift=2pt},
        xlabel style={yshift=3pt},
        every axis label/.append style={font=\scriptsize},
        every axis plot/.append style={
            thick,
            line cap=round,
            line join=round,
        },
    ]

    \nextgroupplot[
        ylabel={Activity (Bq/mL)},
        scaled y ticks=base 10:0,
        legend style={font=\tiny},
        legend image post style={line cap=round, line join=round},
        legend image post style={xscale=0.7},
        legend cell align=left,
        xticklabels={},
        ylabel style={yshift=-1pt},
    ]

    \draw[dashed, Thistle]
        (axis cs:87,\pgfkeysvalueof{/pgfplots/ymin})
        -- (axis cs:87,\pgfkeysvalueof{/pgfplots/ymax});

    \addplot[ref_line] table[x=x, y=ref, col sep=space]{figures/realdata_profiles/profiles_realdata_activity.dat};
    \addlegendentry{Reference}

    \addplot[mlaa_line] table[x=x, y=mlaa, col sep=space]{figures/realdata_profiles/profiles_realdata_activity.dat};
    \addlegendentry{MLAA}

    \addplot[unet_line] table[x=x, y=unet, col sep=space]{figures/realdata_profiles/profiles_realdata_activity.dat};
    \addlegendentry{MLAA-UNet}

    \addplot[dpstof_line] table[x=x, y=jraadps, col sep=space]{figures/realdata_profiles/profiles_realdata_activity.dat};
    \addlegendentry{JRAA-DPS}

    \addplot[dpstofosem_line] table[x=x, y=jraadpsosem, col sep=space]{figures/realdata_profiles/profiles_realdata_activity.dat};
    \addlegendentry{JRAA-DPS + 1 \gls{OSEM} it.}

    \nextgroupplot[
        ylabel={Attenuation (1/mm)},
        xlabel={Pixel index},
        scaled y ticks=base 10:0,
    ]

    \draw[dashed, Thistle]
        (axis cs:87,\pgfkeysvalueof{/pgfplots/ymin})
        -- (axis cs:87,\pgfkeysvalueof{/pgfplots/ymax})
        node[pos=0.60, anchor=south east, xshift=3pt, rotate=90, text=Thistle, font=\tiny] {tumor center};

    \addplot[ref_line] table[x=x, y=ref, col sep=space]{figures/realdata_profiles/profiles_realdata_attenuation.dat};
    \addplot[mlaa_line] table[x=x, y=mlaa, col sep=space]{figures/realdata_profiles/profiles_realdata_attenuation.dat};
    \addplot[unet_line] table[x=x, y=unet, col sep=space]{figures/realdata_profiles/profiles_realdata_attenuation.dat};
    \addplot[dpstof_line] table[x=x, y=jraadps, col sep=space]{figures/realdata_profiles/profiles_realdata_attenuation.dat};
    \addplot[dpstofosem_line, dashed, line cap=butt] table[x=x, y=jraadpsosem, col sep=space]{figures/realdata_profiles/profiles_realdata_attenuation.dat};

    \end{groupplot}
\end{tikzpicture}
	\caption{\textit{Experiment 2}--Activity and attenuation profiles along a tumor (cf. green line in Figure~\ref{subfig:realdata-ref}). Reconstructions from Biograph mMR data (non-\gls{TOF}).}
	\label{fig:profiles_realdata}
\end{figure}

\begin{table}[htbp]
	\footnotesize
	\centering
	\begin{tabular}{ll}
		\toprule
		Biograph mMR: $\bolds$ & proposed method: $\hat{\bolds}(\boldphi)$ \\
		\midrule
		$100 \times\frac{  \|\bolds\|_1}{\|[\boldy - \boldr]_+\|_1 }=50.21\%$ & $100\times\frac{\|\hat{\bolds}(\boldphi)\|_1}{ \|[\boldy - \boldr]_+\|_1  }=36.64\%$ \\
		\bottomrule
	\end{tabular}
	\caption{
		\textit{Experiment 2}---Estimated \gls{SF} with Biograph mMR system and our proposed  method. $\|\cdot\|_1$ denotes the $\ell^1$ norm, i.e., $\|\boldz\|_1 = \sum_i |z_i|$, $\boldr\in \calY$ is the matrix of expected randoms and  $[\cdot]_+$ denotes the positive part. 
	}
	\label{tab:scatter_frac}
\end{table}

\begin{table}[htbp]
	\footnotesize
	\centering
	\begin{tabular}{lr}
		\toprule
		Method & Reconstruction \\ duration (min) \\
		\midrule
		MLAA (non-TOF) & 22 \\
		MLAA-UNet (non-TOF) & 23 \\
		JRAA-DPS (non-TOF) & 77 \\
		\bottomrule
	\end{tabular}
	\caption{\textit{Experiment 2}---Biograph mMR reconstruction duration.}
	\label{tab:recon-time-mmr}
\end{table}

The reference \gls{OSEM} \gls{PET} reconstruction, the reference \gls{MR}-derived attenuation map and their reconstructions using \gls{MLAA}, \gls{MLAA}-UNet and  \gls{JRAA}-\gls{DPS} with joint scatter estimation are shown in Figure \ref{fig:recon-realdata}. While \gls{MLAA} suffers from strong cross-talk which hinders the task of \gls{MLAA}-UNet to recover an accurate attenuation map (cf. Figures \ref{subfig:realdata-mlaa} and \ref{subfig:realdata-unet}), the \gls{MLAA}-UNet activity reconstruction achieves the highest image quality metrics.

\Gls{JRAA}-\gls{DPS} is able to reconstruct both the activity and the attenuation (cf. Figure \ref{subfig:realdata-ours_notof}). However, the method overestimates the activity in lower uptake areas, such as the liver, which degrades the metrics as compared with \gls{MLAA}-UNet. This is caused by an underestimation of the jointly estimated scatter (cf. Table~\ref{tab:scatter_frac}). An evaluation of the influence of the initial \gls{SF} on the joint scatter estimation and the reconstructed images is proposed in appendix.

\Gls{JRAA}-\gls{DPS} recovers all three tumors in the displayed slice, although their magnitude is underestimated. This is can be improved by performing a single \gls{OSEM} iteration (cf. Figures~\ref{subfig:realdata-oursosem_notof} and \ref{fig:profiles_realdata}).

Despite \gls{MLAA}-UNet achieving the best metrics, it should be noted that it is artificially favored for the activity, since both the reference activity image and the \gls{MLAA}-UNet reconstruction are obtained using the same \gls{OSEM} algorithm with the same system's scatter estimate (but with different attenuation inputs). Additionally, \gls{MLAA} and \gls{MLAA}-UNet used the \gls{MR} image support for the attenuation. In contrast, \gls{JRAA}-\gls{DPS} did not use any \gls{MR} prior information.

Reconstruction with \gls{JRAA}-\gls{DPS} takes approximately 1 h 20 min, compared with about 20 min for \gls{MLAA} and \gls{MLAA}-UNet (cf. Table~\ref{tab:recon-time-mmr}). These durations are longer than those reported in \emph{Experiment 1} (cf. Table~\ref{tab:recon-time-test}), primarily due to the larger scanner geometry. In addition, \gls{JRAA}-\gls{DPS} incorporates a joint scatter estimation strategy in this setting, which introduces additional computational overhead. Nevertheless, \gls{JRAA}-\gls{DPS} is only about 3.5 times slower than \gls{MLAA} and \gls{MLAA}-UNet (as opposed to 14 times for simulated data with simpler geometry). This suggests that the proposed method scales reasonably well with increasing problem dimensionality.

	\section{Discussion}\label{sec:discussion}

This paper introduces a framework for \gls{3D} \gls{CT}-less \gls{PET} reconstruction based on \gls{JRAA} using \gls{DPS}. We first simulated data by projecting the reference activity images and attenuation maps. We then evaluated \gls{JRAA}-\gls{DPS} against two other approaches, the classic \gls{MLAA} and \gls{MLAA}-UNet, a post-processing method to predict a pseudoattenuation map from \gls{MLAA} reconstructions. \Gls{JRAA}-\gls{DPS} outperformed the comparison methods with regards to the selected metrics. It is also able to reconstruct non-\gls{TOF} data in higher count settings. In particular, it reconstructed clinical non-\gls{TOF} data acquired with a Siemens Biograph mMR scanner, recovering meaningful structures such as tumors. Overall, \gls{JRAA}-\gls{DPS} reconstructs accurate, noiseless \gls{3D} activity images. 

A key advantage of {JRAA}-\gls{DPS} is that the learned prior is agnostic to scanner geometry, count level, and the presence of \gls{TOF} information, as these factors are explicitly modeled through the forward operator. The framework does not require supervised training to accommodate variations in geometry, noise level, or the absence of \gls{TOF} information. In addition, a joint scatter estimation strategy can be integrated into the framework at sampling time, without requiring additional training. Therefore, a single trained \gls{DM} can be applied to reconstruct data with diverse characteristics (e.g., different count levels, \gls{TOF} or non-\gls{TOF} acquisitions, and varying \glspl{SF}).

However, the diffusion prior is learned from a specific image distribution and may be sensitive to domain shifts not captured by the forward model, such as differences in radiopharmaceutical uptake patterns. In this work, the \gls{DM} was trained and evaluated using data from a single tracer (i.e., \gls{FDG}). Extending the approach to other imaging conditions may therefore require additional training. This limitation could be addressed by training the \gls{DM} on multi-tracer data (e.g., with a tracer-conditioning mechanism), or by developing tracer-specific models through targeted training or fine-tuning.

The joint scatter estimation strategy we used was able to partially mitigate the impact of scatter on reconstructions, but tended to underestimate the scatter contribution, which led to overestimation of the activity in low-uptake regions. In the end, on the real patient dataset, \gls{MLAA}-UNet yielded a higher-quality activity reconstruction than \gls{JRAA}-\gls{DPS} on the clinical data. However, the metrics used for comparison favored \gls{MLAA}-UNet as they were computed with a reference image obtained with a similar method, as noted in Section~\ref{sec:eval_patient}. Additionally, \gls{MLAA}-UNet relies on the scanner-derived scatter estimates and the reference attenuation support, which are both obtained from the attenuation map. Conversely, \gls{JRAA}-\gls{DPS} reconstructs both the activity and the attenuation without relying on any attenuation information, which is a key advantage of the method. More advanced strategies could be applied during joint scatter estimation to improve the quality of the scatter estimate. For instance, the magnitude of the scatter contribution could be constrained based on an expected \gls{SF}, which would mitigate the scatter underestimation.

Generative models are prone to hallucination, meaning they produce features that are not supported by the measurements. In our experiments, these were especially noticeable in the reconstructed attenuation maps: examples include missing bone detail, poorly recovered small air-filled regions, incoherent structures in the lungs, and shifted or extended arms. Although these artifacts did not appear to substantially affect the activity reconstruction in our experiments, especially when they were confined to peripheral or out-of-\gls{FOV} regions, they remain a limitation of the method. However, such artifacts had little effect on the reconstructed activity images in our tests. \Gls{JRAA}-\gls{DPS} was able to recover high-quality activity images, even without \gls{TOF} information, provided the count level was sufficient. In practice, small errors in the attenuation map may not significantly impact the reconstructed activity image, especially given the limited spatial resolution of \gls{PET} systems. The attenuation map obtained in \gls{PET}/\gls{MR} systems is prone to errors (e.g., missing bones, simplified structures) due to the challenges of deriving accurate attenuation information from \gls{MR} images, yet is considered clinically acceptable \cite{catana2020attenuation}. In addition, activity--attenuation cross-talk is reduced when using \gls{TOF} information, which improves robustness to attenuation map errors \cite{mehranian2015impact}. Nonetheless, quantifying the frequency, magnitude and clinical impact of such hallucinations on activity reconstructions is important to assess the clinical viability of diffusion-based reconstruction methods such as \gls{JRAA}-\gls{DPS}.

In \gls{TOF} \gls{PET}, attenuation information is theoretically identifiable from the emission data \cite{defrise2012time}, and the diffusion prior mainly acts as a regularizer. In contrast, attenuation information from non-\gls{TOF} data is more ill-posed, and the reconstructed attenuation map should be interpreted as a solution that is jointly consistent with both the measured data and the learned distribution of plausible $\boldlambda$--$\boldmu$ pairs. Consequently, the prior plays a larger role in non-\gls{TOF} reconstructions, which is reflected by the increased variability observed in Fig. \ref{fig:onetomany}. This stronger dependence on the learned prior may also contributed to hallucinated activity and attenuation outside the object support. In addition, inaccuracies in the jointly estimated scatter may lead to overestimation of activity in low-uptake regions, as observed in the mMR study.

An important limitation of \gls{JRAA}-\gls{DPS} concerns the reconstruction duration. Sampling from \glspl{DM} is known to be time-consuming, particularly when using the standard \gls{DDPM}, and reconstructing a sample of simulated data with \gls{JRAA}-\gls{DPS} currently takes around 30 minutes. Employing the \glsxtrfull{DDIM} with its accelerated inference variant could significantly reduce reconstruction time \cite{song2020denoising}. Additionally, the computation of the gradient of the data fidelity term can also be accelerated \cite{chung2023decomposed,daras2024survey}. However, comparative evaluation would be necessary to assess the impact of this acceleration on reconstruction accuracy.

Finally, \gls{CT}-less \gls{PET} reconstruction methods are also attractive for respiratory motion compensation, since they can eliminate the need for gated-\gls{CT} acquisitions. Previous studies have investigated joint \gls{PET} reconstruction and motion estimation using a single, potentially mismatched, attenuation map \cite{Bousse2016a,Bousse2016b}. Building on these works, \citeauthor{elhamiasl2025joint} proposed two methods for the joint estimation of activity, attenuation and motion \cite{elhamiasl2025joint}. Such strategies could be integrated with our proposed \gls{JRAA}-\gls{DPS} framework to enable diffusion-based motion-compensated \gls{CT}-less \gls{PET} reconstruction.

	\section{Conclusion}\label{sec:conclusion}

In this work, we introduce a novel method for \gls{3D} \gls{CT}-less reconstruction of \gls{PET}. This approach reconstructs accurate and noise-free \gls{PET} images without attenuation information. The framework relies on a prior consisting of a \gls{DM} trained to generate activity-attenuation image pairs. Additionally, we used a  \gls{WDM} to process the large \gls{3D} volumes. Overall, this method enables radiation dose reduction by eliminating the need for a \gls{CT} scan for attenuation correction, and remains effective even in \gls{LC} settings. Coupled with a joint scatter estimation strategy, \gls{JRAA}-\gls{DPS} reconstructs real non-\gls{TOF} clinical data, demonstrating its potential for clinical application. While these results are promising, the current joint scatter estimation tends to underestimate scatter, leading to increased uptake in low-activity areas. In addition, the framework can be affected by hallucinations due to the generative nature of the prior. Nevertheless, \gls{JRAA}-\gls{DPS} remains a promising approach to generate high-quality \gls{PET} images from low-dose or non-\gls{TOF} \gls{PET} data without attenuation information. Future work will focus on improving joint scatter estimation, exploring sampling acceleration methods to reduce reconstruction duration and respiratory motion-compensated \gls{JRAA}.
	
	\appendix
	
	\section*{Impact of the Initialization for Joint Scatter Estimation}

In this section we evaluate the impact of the joint scatter estimation i.e., the scatter update \eqref{eq:scatter_update} used in Algorithm~\ref{algo:dps}, for \textit{Experiment 2} (non-\gls{TOF}).

The joint scatter estimation module is used for the reconstruction of the clinical data in \emph{Experiment 2}. For joint scatter estimation, the scatter term defined in \eqref{eq:scatter_model} is initialized as follows:
\begin{itemize}
    \item compute a first initialization $\bolds^{\mathrm{init}}$ by smoothing the sinogram image obtained after subtracting the randoms estimate from the raw data, and then apply a lower threshold to remove the background,
    \item fit the Gaussian scatter model on this first initialization, i.e., by solving 
    \begin{equation}
    	\text{find $\boldphi$ s.t. } \hat{\bolds}(\boldphi) \approx \bolds^{\mathrm{init}}  \, ,
    \end{equation}
    and
    \item adjust the scaling factors $a_{\nu_i, \varphi_i}$ in \eqref{eq:scatter_model} to correspond to a given initial \gls{SF} (e.g., 25\%, 50\% or 75\%).
\end{itemize}
This initial estimate is subsequently refined during reconstruction using the joint scatter estimation update \eqref{eq:scatter_update}. 
To evaluate the influence of the initial \gls{SF} on the reconstructed images, we compared two strategies: (i) using the initialized scatter directly as a fixed proxy for the scatter distribution  (without refinement step \eqref{eq:scatter_update}), and (ii) jointly refining this initial estimate during reconstruction (with refinement step \eqref{eq:scatter_update}). For reference, we also reconstructed the data using the scanner-derived scatter estimate.

Table~\ref{tab:scatter} reports the final \glspl{SF} (at the end of Algorithm~\ref{algo:dps}). The results show that the jointly estimated scatter is robust to the scaling of the initialization. For all tested initializations, the final \gls{SF} converges to approximately 36.5\%, and the reconstructed activity images exhibit very similar quantitative metrics and visual appearance (cf. Figure~\ref{fig:scatter}). The scanner-derived scatter estimate corresponds to a substantially higher \gls{SF} (50.2\%). This discrepancy suggests that the proposed scatter estimation module underestimates the total scatter contribution, with part of the scatter signal being absorbed into the reconstructed activity.

\begin{table}[h]
	\centering
	\footnotesize
	\begin{tabular}{llr}
		\toprule
		Method & Scatter setting & Final \gls{SF} (\%) \\
		\midrule
		
		\makecell[l]{Reference (OSEM with\\ MR-derived attenuation)} 
		& Scanner-derived & 50.21 \\
		
		\thinmidrule
		
		\multirow[c]{4}{*}{No scatter refinement} 
		& Scanner-derived & 50.21 \\
		& Init. 25\% & 25.00 \\
		& Init. 50\% & 50.00 \\
		& Init. 75\% & 75.00 \\
		
		\thinmidrule
		
		\multirow[c]{3}{*}{With scatter refinement} 
		& Init. 25\% & 36.40 \\
		& Init. 50\% & 36.64 \\
		& Init. 75\% & 37.41 \\
		
		\bottomrule
	\end{tabular}
	
	\caption{
		\textit{Experiment 2}---Comparison of \gls{JRAA}-\gls{DPS} scatter estimates with various scatter settings.
	} 
	\label{tab:scatter}
\end{table}

\input{figures/fig_scatter}

More advanced regularization strategies could be applied during joint scatter estimation to improve the quality of the scatter estimate. For instance, the magnitude of the scatter contribution could be constrained, for example by including a penalty term in \eqref{eq:scatter_update} to prevent the \gls{SF} to deviate from a reference value that is computed from patient-specific parameters (e.g., size and weight).

	\section*{Acknowledgment}
	
	All authors declare that they have no known conflicts of interest in terms of competing financial interests or personal relationships that could have an influence or are relevant to the work reported in this article.
	
	\AtNextBibliography{\scriptsize} 
	\printbibliography

\end{document}